\documentclass[usenatbib]{mn2e}

\def\Zsol{\hbox{Z$_{\odot}$}}
\def\Msol{\hbox{M$_{\odot}$}}

\usepackage{times}
\usepackage{epsfig,natbib,url,graphicx}
\usepackage{lscape,amsmath,amssymb}
\newcommand{\hi}{H\,{\sc i}}
\newcommand{\hii}{H~{\sc ii}}
\newcommand{\hei}{He~{\sc i}}
\newcommand{\heii}{He~{\sc ii}}
\newcommand{\kms}{km\,s$^{-1}$}
\newcommand{\eld}{$N_{\rm e}$}

\newcommand{\elt}{$T_{\rm e}$}

\newcommand{\Sp}{S$^+$}
\newcommand{\Spp}{S$^{2+}$}

\newcommand{\op}{O$^+$}

\newcommand{\opp}{O$^{2+}$}

\newcommand{\np}{N$^+$}

\newcommand{\nepp}{Ne$^{2+}$}

\newcommand{\foiii}{[O~{\sc iii}]}

\newcommand{\foii}{[O~{\sc ii}]}
\newcommand{\fsii}{[S~{\sc ii}]}
\newcommand{\fsiii}{[S~{\sc iii}]}

\newcommand{\fnii}{[N~{\sc ii}]}

\newcommand{\fneiii}{[Ne~{\sc iii}]}

\newcommand{\ffeiii}{[Fe~{\sc iii}]}

\newcommand{\ciii}{C~{\sc iii}}

\newcommand{\hp}{H$^+$}

\newcommand{\ha}{H$\alpha$}
\newcommand{\hb}{H$\beta$}
\newcommand{\hg}{H$\gamma$}
\newcommand{\hd}{H$\delta$}

\title[A 3D Chemodynamical Study of UM~448]{The merging dwarf galaxy UM 448: chemodynamics of the ionized gas from VLT integral field spectroscopy}
\author[B. L. James et al.]{B.~L. James$^{1}$\thanks{E-mail: bjames@stsci.edu},
Y.~G.~Tsamis$^{2}$,
M.~J.~Barlow$^{3}$,
J.~R.~Walsh$^{2}$,
M.~S.~Westmoquette$^{2}$\\
$^{1}$Space Telescope Science Institute, Baltimore, MD 21218\\
$^{2}$European Southern Observatory, Karl-Schwarzschild Strasse 2, D-85748 Garching bei
M\"unchen, Germany\\
$^{3}$Department of Physics and Astronomy, University College London, Gower Street, London, WC1E 6BT}

\begin{document}

\date{Accepted ?. Received April 2012}

\pagerange{\pageref{firstpage}--\pageref{lastpage}} \pubyear{2002}

\maketitle

\label{firstpage}

\begin{abstract}
Using VLT/FLAMES optical integral field unit observations, we present a detailed study of UM~448, a nearby Blue Compact Galaxy (BCG) previously reported to have an anomalously high N/O abundance ratio.  NTT/SuSI2 images reveal a morphology suggestive of a merger of two systems of contrasting colour, whilst our \ha\, emission maps resolve UM~448 into three separate regions that do not coincide with the stellar continuum peaks. UM~448 exhibits complex emission line profiles, with most lines consisting of a narrow (FWHM$\lesssim$100~\kms), central component, an underlying broad component (FWHM$\sim$150--300~\kms) and a third, narrow blue-shifted component. Radial velocity maps of all three components show signs of solid body rotation across UM~448, with a projected rotation axis that correlates with the continuum morphology of the galaxy.
A spatially-resolved, chemodynamical analysis based on the \foiii~$\lambda\lambda$4363, 4959, \fnii~$\lambda$6584, \fsii$\lambda\lambda$6716, 6731 and \fneiii~$\lambda$3868 line maps, is presented. Whilst the eastern ÔtailÕ of UM~448 has electron temperatures (\elt) that are typical of BCGs, we find a region within the main body of the galaxy where the narrow and broad \foiii~$\lambda$4363 line components trace temperatures differing by 5000~K and oxygen abundances differing by 0.4~dex. We measure spatially resolved and integrated ionic and elemental abundances for O, N, S and Ne throughout UM~448, and find they do not agree, possibly due the flux-weighting of \elt\, from the integrated spectrum. This has significant implications for abundances derived from long-slit and integrated spectra of star-forming galaxies in the nearby and distant universe. A region of enhanced N/O ratio is indeed found, extended over a $\sim$0.6~kpc$^2$ region within the main body of the galaxy. Contrary to previous studies, however, we do not find evidence for a large Wolf-Rayet population, and conclude that WR stars alone cannot be responsible for producing the observed N/O excess. Instead, the location and disturbed morphology of the N-enriched region suggests that interaction-induced inflow of metal-poor gas may be responsible.
\end{abstract}

\begin{keywords}
galaxies:abundances --
galaxies: individual (UM~448) --
galaxies:dwarf --
galaxies:kinematics and dynamics --
galaxies:interactions --
stars: Wolf-Rayet
\end{keywords}

\section{Introduction}
\label{sec:intro}

Observations of Blue Compact Galaxies (BCGs) in the nearby Universe 
provide a means for studying chemical evolution and star formation (SF) 
processes in chemically un-evolved environments.  Their low metallicities 
make them attractive as analogues to the young `building-block' galaxies
thought to exist in the high-$z$ primordial universe \citep{Searle:1972}.  BCGs are
thought to have experienced only very low-level star formation in the past, but are
currently undergoing bursts of SF in relatively pristine environments, 
ranging from 0.02--0.5 solar metallicity \citep{Kunth:2000}.  Thus
understanding the chemical evolution within BCGs can impact upon our understanding
of primordial galaxies and galaxy evolution in general.

The distribution of nitrogen as a function of metallicity in SF galaxies 
has often been a cause for debate, especially at intermediate 
metallicities (7.6$\leq$ 12+log(O/H) $\leq$8.3).  Within this metallicity range the 
N/O abundance ratio indicates that nitrogen behaves neither as a primary element 
(i.e. N/O independent of O/H) nor as a secondary element (i.e. N/O 
proportional to O/H); instead a large scatter is observed.  In particular, 
several BCGs exist whose N/O abundance ratios are up to three times those
expected for their O/H values, as highlighted by e.g. \citet{Pustilnik:2004}.  
Several of these galaxies have been the focus of our spectroscopic studies
using integral field unit (IFU) observations \citep{James:2009,James:2010}, aimed at
understanding the source of such abundance peculiarities.

\citet{James:2009} performed a detailed abundance
analysis of the N/O-outlier BCG, Mrk~996, whose X-ray properties have been studied by \citet{Georgakakis:2011}. Previous long-slit
observations of this galaxy \citep{Thuan:1996,Pustilnik:2004},  had
shown it to have N/O ratio values at least $\sim$4 times higher than those of
other BCGs of similar metallicity. Using IFU observations, \citet{James:2009} performed a spatially
resolved multi-component emission line analysis of Mrk~996, which
revealed a twenty-fold N/O and N/H enrichment factor in the broad component
(400--500 km/s full width at half maximum -- FWHM) versus the narrow component  line emission. The
narrow component gas on the other hand has a N/O ratio which is typical for the galaxy's
metallicity. In this respect, Mrk~996 is
similar to NGC~5253, a starburst galaxy for which an enhanced N/O ratio has also been
isolated to broad component gas \citep{Lopez-Sanchez:2010}. In Mrk~996, the spatial
morphology of the broad line component is highly correlated with Wolf-Rayet (WR)
wind-feature emission arising from a population of WNL stars, strongly suggesting the
enhanced nitrogen to be the result of chemical enrichment by the N-rich winds of WN-type
WR stars. This supports the postulate of
nitrogen self-enrichment due to WR stars that had been suggested by
\citet{Walsh:1987} and \citet{Pagel:1986} to explain elevated N/O levels.

\citet{James:2010} also presented an integral field spectroscopic
study of the interacting dwarf galaxy UM~420 for which the long-slit spectroscopic study
of \citet{Izotov:1999} had reported log(N/O) $=$ $-$1.08. By
performing a spatially resolved analysis of its physical and chemical 
conditions, \citet{James:2010} found an average ratio applicable to the whole galaxy of log(N/O) $=$
$-$1.64, i.e. 3.6 times lower than previous estimates. The VLT observations of \citet{James:2010} did not
confirm previous detections of WR spectral features in UM~420, highlighting
the difficulties involved in identifying WR stellar populations in distant dwarf galaxies.

In this paper, we present IFU observations of another member of this
subset of potentially N/O-anomalous galaxies, UM~448 (UGC~06665, Mrk~1304). With a
reported factor of $\sim$3 nitrogen excess \citep{Bergvall:2002}, this galaxy
presents an excellent opportunity to further our understanding of peculiar N/O 
ratios in some BCGs and the mechanisms responsible.  UM~448 is a complex system, with a
disturbed morphology possibly due to a merger event \citep{Dopita:2002}. Thus the need for
spatially resolved spectroscopy is two-fold: firstly in determining the range of physical
properties and chemical compositions that are present, and secondly, in 
spatially relating any chemical anomalies to environmental factors (e.g. WR populations).


UM~448 is a starburst galaxy \citep{Keel:1992,Kewley:2001} consisting of
various star forming knots, diffuse \ha\ emission in the north, and a SE tidal extension
(\citet{Dopita:2002}). Broad \heii\ $\lambda$4686 emission was detected by
\citet{Masegosa:1991} and subsequently by \citet{Guseva:2000}, who reported an atypical
weakness in this line in comparison with other WR-signature lines.  As with Mrk~996 and
NGC~5253, a WR population may well be the source of N-enhancement within this BCG.

In this paper we present integral field spectroscopy (IFS) observations obtained with the
VLT-FLAMES IFU. Crucially, this instrument is sufficiently blue-sensitive to detect the
\foii\ $\lambda$3727 doublet line, meaning that we are not required to adopt uncertain
ionisation correction factors to determine oxygen abundances. These data afford us a new
spatiokinematic 3-D view of UM 448, with the spatial and spectral resolution to
undertake a full multi-velocity component analysis. We adopt a distance
of 75.9\,Mpc for UM~448 from the velocity measured
here of $+$5579$\pm5$\, km~s$^{-1}$ ($z=0.018660\pm1.6\times10^{-5}$), with a 
Hubble constant of H$_o=$73.5\,km~s$^{-1}$~Mpc$^{-1}$ 
\citep{Bernardis:2008}.

\begin{figure*}
\includegraphics[scale=.60]{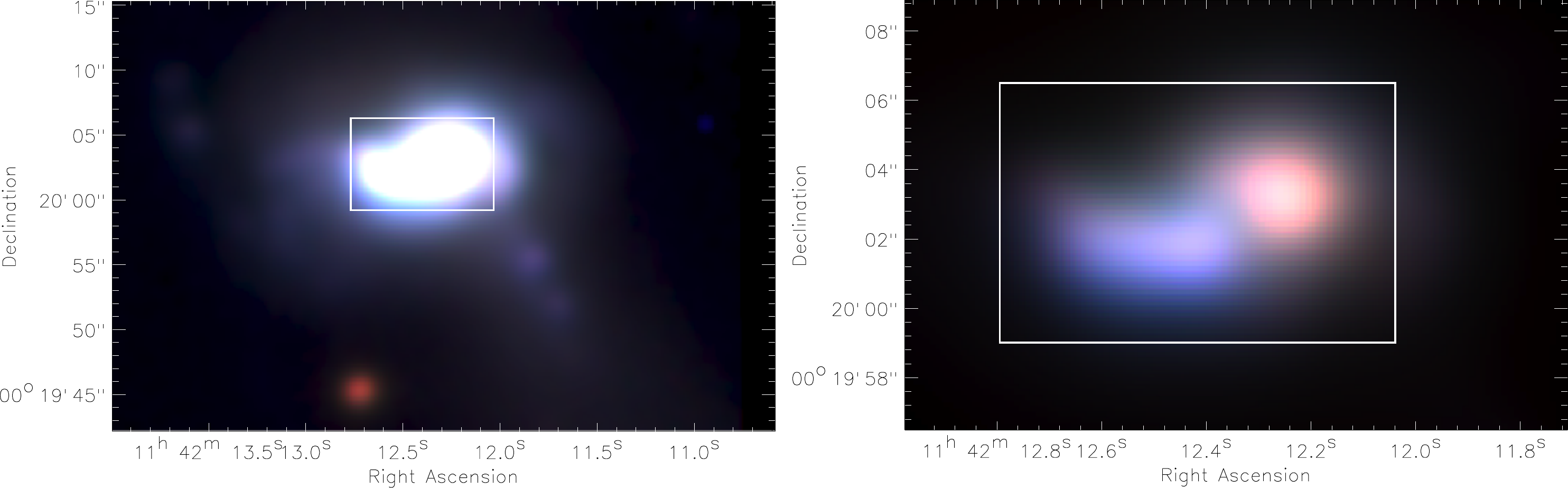}
\caption{SuSI2 (Superb-Seeing Imager, installed on the ESO 3.5m New Technology Telescope
(NTT) at La Silla Observatory, Chile) colour-composite images of UM~448 (red, $R$-band;
green, $V$-band; blue $B$-band), with the 11.5\arcsec $\times$7.3\arcsec\, FLAMES IFU
aperture
overlaid.  Each image has been scaled to enhance different features within UM~448.  The
left-panel shows two large knots of continuum or ionised gas emission that extend away
from the galaxy in a south-westerly direction, outside the IFU FoV.  The right-panel shows
the two distinct portions of the galaxy, a blue eastern tail and a reddish knot of stellar
continuum.  SuSI2 images were obtained from the ESO archive, observed as part of programme 078.B-0358(A).}
\label{fig:um448_susi}
\end{figure*}

\begin{table*}
\begin{center}
\begin{footnotesize}
\caption{FLAMES IFU observing log}
\begin{tabular}{ccccccc}
Date &Grism & Wavelength Range   &Exp. time & Avg. Airmass & FWHM seeing & Standard Star\\
 & & (\AA) & (s) & & (arcsec) & \\
\hline\hline
24/05/2009 & LR1 & 3620--4081 & 3${\times}$233 & 1.11  & 0.85 & LTT7379 \\
.. & LR2 & 3964--4567 & 3$\times$300 & 1.12 & 0.87 & LTT7379\\
.. & LR3 & 4501--5078 & 3$\times$230 & 2.24 & 0.80 & LTT7379\\
18/06/2009 & LR1 & 3620--4081 & 3${\times}$233 & 1.12  & 0.47 & Feige110 \\
.. & LR2 & 3964--4567 & 3$\times$300 & 1.13 & 0.42 & Feige110\\
.. & LR3 & 4501--5078 & 3$\times$230 & 1.16 & 0.50 & Feige110\\
28/07/2009 & LR6 & 6438--7184 & 3$\times$400 & 1.75 & 0.47 & EG21\\
26/01/2010 & LR1 & 3620--4081 & 3${\times}$233 & 1.15  & 0.54 & LTT4364\\
.. & LR2 & 3964--4567 & 3$\times$300 & 1.13 & 0.53 & LTT4364\\
.. & LR3 & 4501--5078 & 3$\times$230 & 1.11 & 0.63 & LTT4364\\
16/03/2010 & LR1 & 3620--4081 & 3${\times}$233 & 1.10  & 0.57 & LTT4364 \\
.. & LR2 & 3964--4567 & 3$\times$300 & 1.11 & 0.50 & LTT4364\\
.. & LR3 & 4501--5078 & 3$\times$230 & 1.12 & 0.56 & LTT4364\\
08/07/2010 & LR6 & 6438--7184 & 8$\times$600 & 1.63 & 0.77 & Feige110\\
09/07/2010 & LR1 & 3620--4081 & 3${\times}$233 & 1.50  & 0.77 & Feige110\\
.. & LR2 & 3964--4567 & 3$\times$300 & 1.60 & 0.72 & Feige110\\
.. & LR3 & 4501--5078 & 3$\times$230 & 1.74 & 0.66 & Feige110\\
\hline\label{tab:obs}
\end{tabular}
\end{footnotesize}
\end{center}
\end{table*}

\section{Observations and data reduction}
\label{sec:obs}

Observations of UM~448 were obtained with the Fibre Large Array Multi Element Spectrograph, FLAMES \citep{Pasquini:2002} at Kueyen, Telescope Unit 2 of 
the 8.2\,m VLT at ESO's observatory on Paranal, in service-mode on dates specified in 
Table~\ref{tab:obs}. Observations were made with the Argus Integral Field
Unit (IFU), with a field-of-view (FoV) of 11\farcs5$\times$7\farcs3 and a 
sampling of 0\farcs52/lens.  In addition to science fibres, Argus has 15
sky-dedicated fibres that simultaneously observe the sky and that were 
arranged to surround the IFU FoV at distances of $\sim$35--58\arcsec.  The positioning of the IFU aperture for 
UM~448 is shown overlaid on SuSI2 images\footnote{SuSI2 images were obtained from the ESO archive, with total exposure times of 1200, 1200, 1800 and 2700\,s, average airmasses of 1.34, 1.28, 1.22 and 1.17 and image qualities of 0.75\arcsec, 0.93\arcsec, 0.87\arcsec\, and 0.56\arcsec\, FWHM for the $R$-, $V$-, $B$- and $U$-band images, respectively.  The average image quality of the images is $\sim$0.78\arcsec\, FWHM.  WCS information was assigned via position matching with GSC2.  Other than aligning and median-combining the separate frames using IRAF's \textsc{imalign} and \textsc{imcombine}, respectively, no additional reduction was performed on the images.} in
Fig.~\ref{fig:um448_susi}.

Four different low-resolution (LR) gratings were utilised: LR1
($\lambda$$\lambda$3620-4081, 24.9$\pm$0.3~\kms\, FWHM resolution), LR2 
($\lambda$$\lambda$3964-4567, 24.7$\pm$0.1~\kms), LR3
($\lambda$$\lambda$4501-5078, 24.6$\pm$0.3~\kms) and LR6 
($\lambda$$\lambda$6438-7184, 21.9$\pm$0.4~\kms).  This 
enabled us to cover all the important emission lines needed for an optical 
abundance analysis, as detailed in Section~\ref{sec:mapping}.  The 
photometric conditions, airmass and exposure time for each data set are 
detailed in Table~\ref{tab:obs}. In addition to the science frames, spectrophotometric standard star observations (Table~\ref{tab:obs}),
as well as continuum and ThAr arc lamp exposures were obtained.

The data were reduced using the girBLDRS pipeline (Blecha \& Simond 2004)\footnote{The girBLDRS pipeline is provided by the Geneva Observatory 
http://girbldrs.sourceforge.net} in
a process which included bias removal, localisation of fibres on the
flat-field exposures, extraction of individual fibres, wavelength
calibration and rebinning of the ThAr lamp exposures, and the full
processing of science frames which resulted in flat-fielded,
wavelength-rebinned spectra. The pipeline recipes `biasMast', `locMast',
`wcalMast', and `extract' were used \citep[cf.][]{Tsamis:2008}.
The frames were then averaged using the `imcombine' task of IRAF\footnote{IRAF is
distributed by the National
Optical Astronomy Observatory, which is operated by the Association of 
Universities for Research in Astronomy} which also
performed the cosmic ray rejection. The flux calibration was performed within IRAF using
the tasks \textsc{calibrate} and
\textsc{standard}.  Spectra of the spectrophotometric stars quoted in
Table~\ref{tab:obs} were individually
extracted with girBLDRS and the spaxels containing the stellar emission were summed up
to form a 1D spectrum. The sensitivity function was determined using 
\textsc{sensfunc} and this was subsequently applied to the combined
UM~448 science exposures. The sky subtraction was performed by averaging the 
spectra recorded by the sky fibres and subtracting this spectrum from that 
of each spaxel in the IFU. Custom-made scripts were then used to convert 
the row by row stacked, processed CCD spectra to data cubes.  This resulted 
in one science cube per grating, i.e. four science cubes for the source
galaxy.


Observations of an object's spectrum through the Earth's atmosphere
are subject to refraction as a function of wavelength, known
as differential atmospheric refraction (DAR). The direction of DAR
is along the parallactic angle at which the observation is made.
Following the method described in James et al. (2009), each reduced data
cube was corrected for this effect using the algorithm created by \citet{Walsh:1990} ; this proceedure calculates the fractional pixel shifts 
for each monochromatic slice of the cube relative to a fiducial wavelength 
(i.e. a strong emission line), shifts each slice with respect to the 
orientation of the slit on the sky and the parallactic angle and 
re-combines the DAR-corrected cube.  All pixel shifts were $\lesssim1$~pixel in size, indicating uncertainties due to extinction or temperature gradients are negligible.

\section{Mapping of Line Fluxes and Kinematics}
\label{sec:mapping}
\begin{table}
\begin{center}
\begin{scriptsize}
\begin{tabular}{|l|cc|}
\hline
	&	F$_{\lambda}$			&	I$_{\lambda}$									\\
	\hline\hline
\foii\, $\lambda$3727	&	98.02	$\pm$		1.18	&	125.40	$\pm$	6.20	\\
\foii\, $\lambda$3729	&	135.65	$\pm$		1.43	&	173.55	$\pm$	8.52	\\
\fneiii\, $\lambda$3868	&	23.41	$\pm$		1.71	&	29.24	$\pm$	2.54	\\
H8+\hei\, $\lambda$3888	&	15.47	$\pm$		2.24	&	19.26	$\pm$	2.93	\\
\hd\, 	&	21.57	$\pm$		1.21	&	25.70	$\pm$	1.85	\\
\hg\,	&	43.33	$\pm$		0.83	&	48.96	$\pm$	2.31	\\
\foiii\, $\lambda$4363	&	1.90	$\pm$		0.19	&	2.14	$\pm$	0.23	\\
\hei\, $\lambda$4471	&	4.50	$\pm$		0.49	&	4.94	$\pm$	0.57	\\
\hb\,	&	100.00	$\pm$		0.99	&	100.00	$\pm$	3.94	\\
\foiii\, $\lambda$4959	&	91.86	$\pm$		1.66	&	89.76	$\pm$	3.71	\\
\fnii\, $\lambda$6548	&	15.83	$\pm$		0.52	&	11.66	$\pm$	0.49	\\
\ha\,	&	407.05	$\pm$		13.27	&	301.47	$\pm$	12.62	\\
\fnii\ $\lambda$6584	&	58.55	$\pm$		0.51	&	42.91	$\pm$	1.17	\\
\hei\, $\lambda$6678	&	2.86	$\pm$		0.06	&	2.07	$\pm$	0.07	\\
\fsii\, $\lambda$6716	&	49.17	$\pm$		0.44	&	35.38	$\pm$	0.94	\\
\fsii\, $\lambda$6731	&	36.07	$\pm$		0.34	&	25.95	$\pm$	0.70	\\
										
c(\hb) &\multicolumn{2}{c|}{		0.42	$\pm$		0.02			}		\\
F(\hb) &\multicolumn{2}{c|}{		24.48	$\pm$		0.17			}		\\

\hline
\end{tabular}

\caption{UM~448 fluxes and de-reddenened line intensities (both relative to F(\hb)=I(\hb)=100) for summed spectra over all of UM~448.  Line fluxes were extinction-corrected using the c(\hb) values shown at the bottom of the table, calculated from the relative \ha, \hb\, and \hg\, fluxes.  The line intensities listed here were used for summed-spectra \elt\, and \eld\, diagnostics and regional ionic abundance calculations listed in Table~\ref{tab:um448_ab}.}
\label{tab:um448_sum}
\end{scriptsize}
\end{center}
\end{table}
%

\begin{figure*}
\includegraphics[scale=.70]{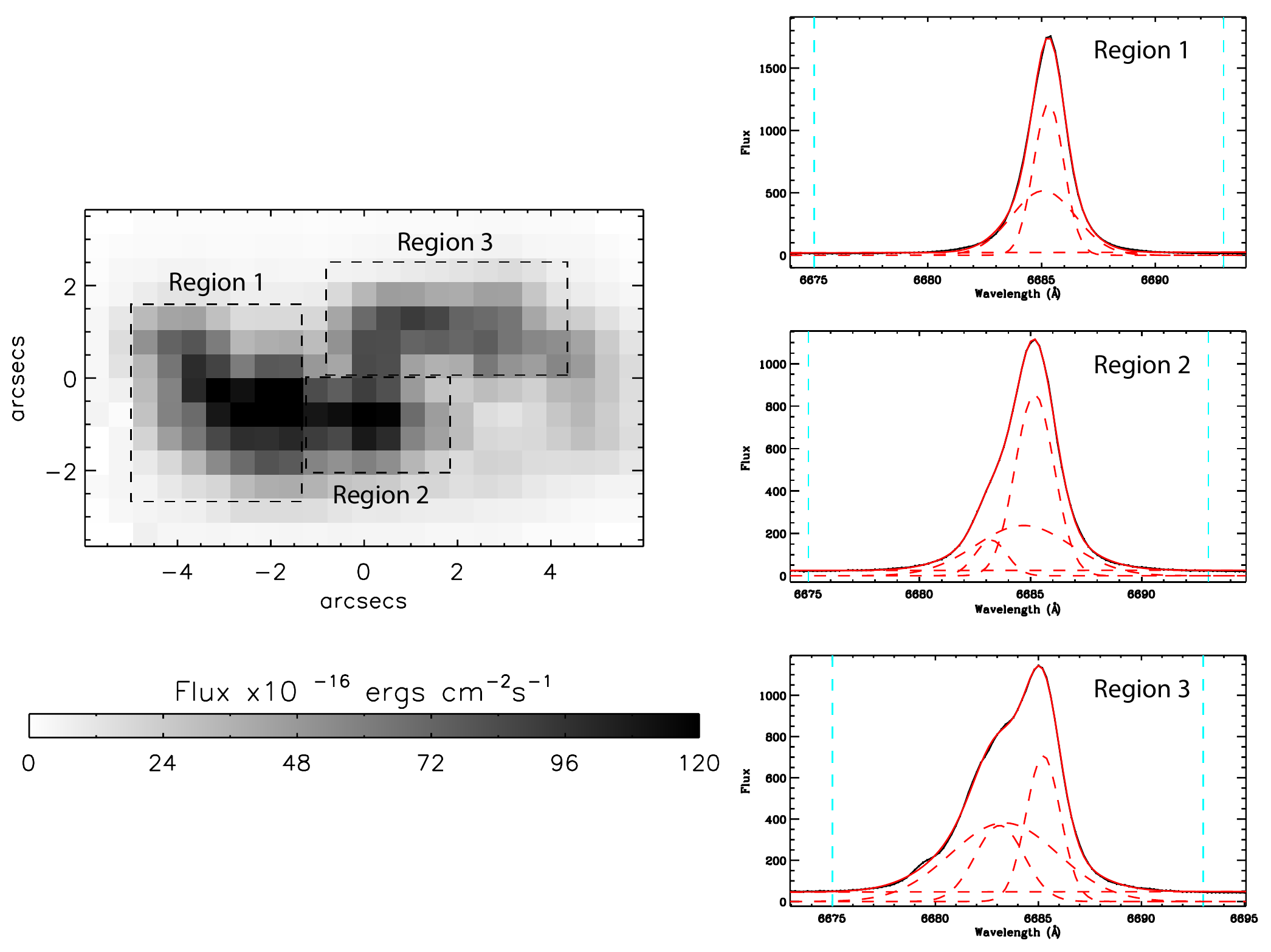}
\caption{\ha\, emission per 0\farcs52$\times$0\farcs52~arcsec$^2$
spaxel. Three regions were defined across UM~448 over which summed spectra were extracted
for analysis. On the right the \ha\ line profile from each summed region is shown,
together with a two-component Gaussian fit. Blue dashed lines demarcate the wavelength range used for summing the flux.  North is up and east is to the left.}
\label{fig:um448_SFR}
\end{figure*}

The distribution of H~{\sc i} Balmer line emission across UM~448 which is indicative of
current massive star formation activity, and the morphology of the system as it appears on
the SuSI2 images, were used to define three regions of roughly similar Balmer line
luminosity. Our analysis of the galaxy is based on properties averaged over each of
these three areas whose boundaries are displayed on the \ha\, map in
Figure~\ref{fig:um448_SFR}. Regions 1 and 2 are both contained in the blue eastern tail of
UM~448 as seen on Fig.~\ref{fig:um448_susi} (right), containing the bulk of Balmer line emission. Region 3
encompasses the interface of the blue eastern tail with the redder western part of the
galaxy, and further contains the peak of the continuum emission in the vicinity of \ha\, (see Fig.\,~12). Whilst each of these regions shows a relatively small variation in
surface brightness, the composite structure of the emission lines can vary considerably
as illustrated by the \ha\, emission
line profiles displayed alongside the monochromatic line map shown in
Figure~\ref{fig:um448_SFR}.

The high spectral resolution and signal-to-noise (S/N) ratio of the data allow the identification of multiple velocity components in the majority of
emission lines seen in the 300 spectra of UM~448 across the IFU aperture.
Following the methods outlined in \citet{James:2009} and \citet{Westmoquette:2007a}, we utilised an automated fitting proceedure
called PAN \citep[Peak ANalysis;][]{Dimeo:2005}, that allowed us to read 
in multiple spectra simultaneously, interactively specify initial 
parameters for a spectral line fit and then sequentially process each 
spectrum, fitting Gaussian profiles accordingly.  The output consists of 
the $\chi^2$ value for the fit, the fit results for the continuum and
each line's flux, centroid and FWHM, and the respective
errors of these quantities which were propagated in all physical 
properties derived afterwards\footnote{The uncertainties output by the pipeline were
used as the starting point of the error analysis; they
also include uncertainties from the flat-fielding and the geometric 
correction of the spectra and not just photon-noise \citep{Blecha:2004}.}.

Even though some low S/N lines may in reality be composed of multiple emission components, it is often not statistically
significant to fit anything more than a single Gaussian. In order to rigorously determine the optimum number of Gaussians required to fit each
observed profile, we followed the likelihood ratio method outlined by \citet[][their Appendix A]{Westmoquette:2011}. S/N ratio
maps were first made for
each emission line, taking the ratio of the integrated intensity
of the line to that of the error-array produced by the FLAMES pipeline, on 
a spaxel-by-spaxel basis. Single or multiple Gaussian profiles were then
fitted to the emission lines, restricting the minimum FWHM to be the instrumental width
(see Section~\ref{sec:obs}). Suitable wavelength limits were defined for
each line and continuum level fit.  Further constraints were
applied when fitting the \fsii\, and \foii\, doublets: the wavelength 
difference between each line in the doublet was taken to be equal to the 
redshifted laboratory value when fitting the velocity component, and their 
FWHMs were set equal to one another.

\begin{figure*}
\includegraphics[scale=.6]{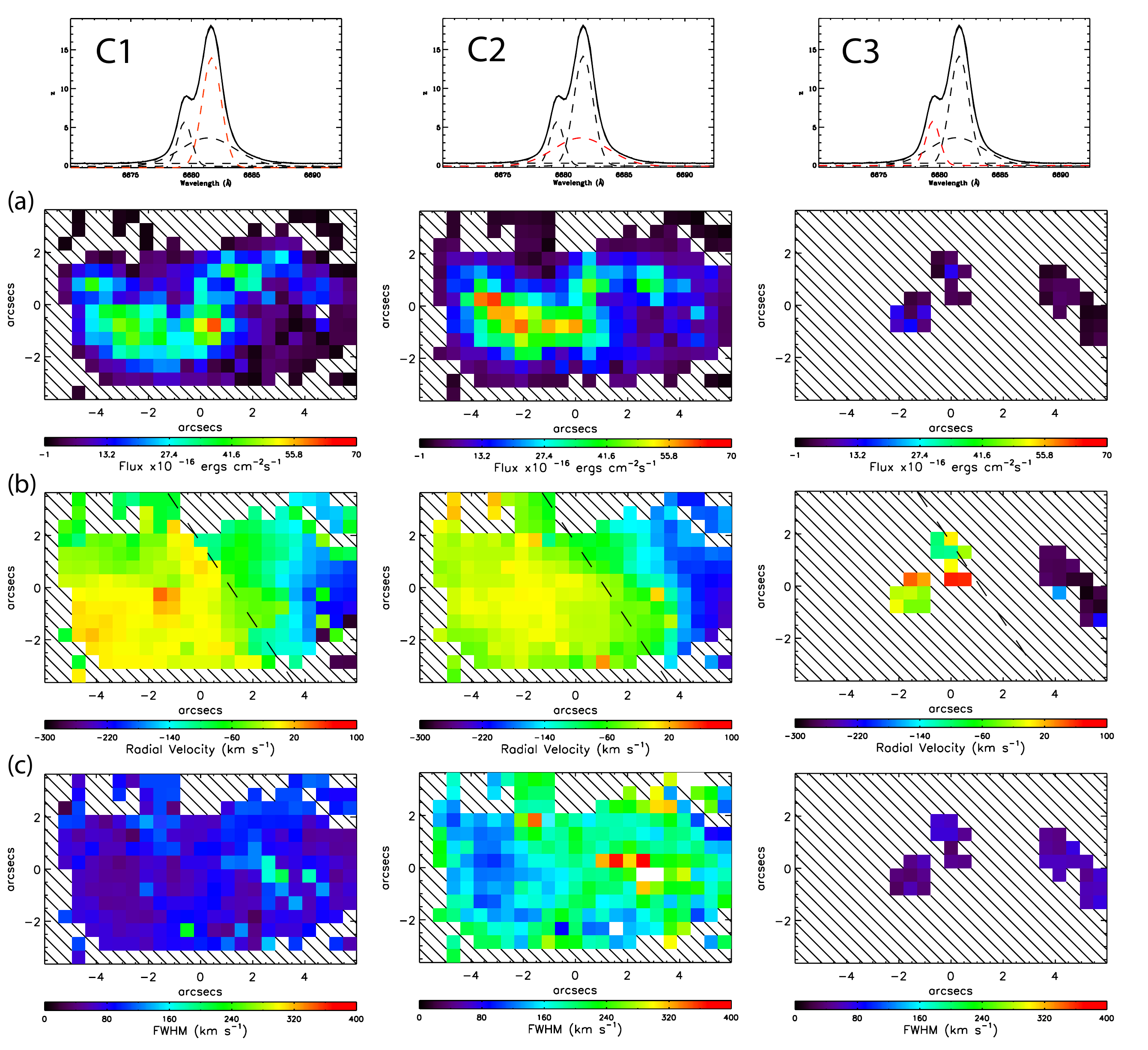}
\caption{Maps of UM~448 in the \ha\, velocity components C1--C3: the top row shows an
example 3-component fit to the \ha\ line and the structure of the velocity component shown in red is displayed below; a) flux in units of
$10^{-16}~ergs~s^{-1}~cm^{-2}~arcsec^{-2}$ per 0\farcs52$\times$0\farcs52 spaxel; b)
radial velocity relative to the heliocentric systemic velocity of $+$5579~\kms (the
dashed line refers to the PA$\sim$36$^\circ$ rotation axis discussed in the text; c) FWHM corrected for the
instrumental PSF. See text for details.  Maps are shown for S/N $\geq$ 10 in the integrated line profile. North is up
and east is to the
left.}
\label{fig:um448_ha}
\end{figure*}

The majority of high S/N ratio line profiles in the spectra of UM~448 are optimally fitted
with a single narrow Gaussian (FWHM$<$100~\kms, component C1 hereafter), a single underlying broad
Gaussian component (FWHM 150-300~\kms, C2), and a third narrow Gaussian which can
appear either red- or blue-shifted with respect to C1 (C3 hereafter).  In consideration of the different `features' of each component (i.e. FWHM and/or velocity), we consider each component as being real, i.e. arising from gas with different physical conditions.
Towards the outer regions of the galaxy, where the S/N ratio decreases, the 
lines are optimally fitted with only C1 and C2 velocity components.  Low 
S/N ratio emission lines (e.g. H8+\hei\, $\lambda$3888) were optimally fitted with 
a single narrow Gaussian profile (C1). Figure~\ref{fig:um448_ha}
shows the complex profile of the \ha\, line, along with the spatial distribution of its constituent velocity components.

\begin{figure}
\includegraphics[scale=1.0]{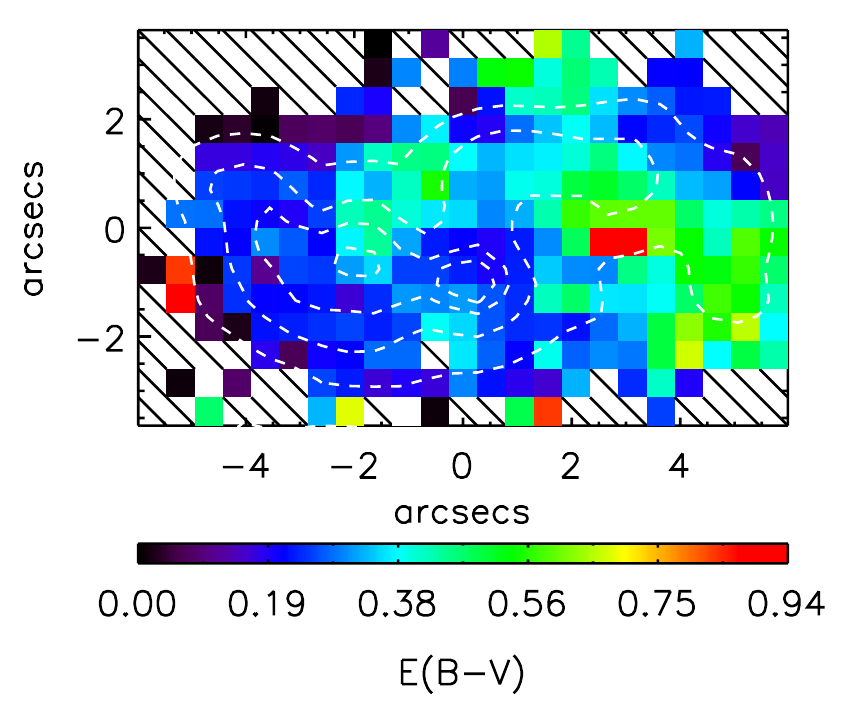}
\caption{Map of E(B-V) for UM~448, derived from the integrated flux of H~{\sc i} emission
lines. Integrated light \ha\, emission line contours are overlaid.  North is up and east is to the left.}
\label{fig:um448_red}
\end{figure}

Tables \ref{tab:um448-1}-\ref{tab:um448-3} list measured FWHMs and 
observed and de-reddened fluxes for the fitted Gaussian components of the
detected emission lines across the three regions of massive star formation defined
in Fig.\,~\ref{fig:um448_SFR}. The fluxes are for spectra summed over each region
and are quoted relative to the flux of the corresponding \hb\, component.
They were corrected for reddening using the galactic reddening law of
\citet{Howarth:1983} using c(\hb) values derived from the \ha/\hb\, and
\hg/\hb\, line ratios of their corresponding components, weighted in a 3:1
ratio \footnote{An exception was made for region 3 where the
low S/N ratio of \hg\ prohibited the robust decomposition into the three
components observed in \ha\, and \hb. Therefore, only the
\ha/\hb\, line ratio was used for the calculation of c(H$\beta$) for this
region.}, respectively, in conjunction with theoretical Case B ratios from
\citet{Storey:1995}. An extinction map was also derived using the the \ha/\hb\, and \hg/\hb\, emission line ratio maps and is shown in Fig.~\ref{fig:um448_red}. This was computed using the integrated line profiles (i.e. C1$+$C2$+$C3) to
obtain a global extinction view of the galaxy. 

A Milky Way reddening of $E(B-V)=$ 0.025~mag in
the direction to UM~448 is indicated by the extinction maps of
\citet{Schlegel:1998}, corresponding to c(\hb)=0.036. Total c(\hb) values applicable to the galaxy and its
individual emission line components in each main aperture-region are listed in
their respective tables (Tables \ref{tab:um448-1}-\ref{tab:um448-3}).

\subsection{\ha\, maps}
The \ha\, surface brightness of UM~448 (Fig.~\ref{fig:um448_ha}a)
displays a horizontal `S'-shaped morphology, extending a total of $\sim10''$ east
to west, almost symmetrically about the FoV center.  The lower arm of
the galaxy extends $\sim5''$ to the east and contains the galaxy's peak \ha\ emission
(corresponding to designated region 1).
Based on our fitting method, the average surface brightness within region 1 can be
de-composed to $\sim1.8\times10^{-14}$
and $\sim2.8\times10^{-14}$~ergs~cm$^{-2}$~s$^{-1}$~arcsec$^{-2}$ for line components C1
and C2 respectively. The upper arm is seen to extend $\sim5''$ in the opposite direction
and displays an overall slightly lower surface brightness of
$\sim1.2\times10^{-14}$~ergs~cm$^{-2}$~s$^{-1}$~arcsec$^{-2}$ for both components.
Component C3 appears in three separate regions; two regions either side of the central
boundary region, i.e. between the two stellar components (see Fig.~\ref{fig:um448_susi}), and a third
region along the most western part of the upper arm. C3 had an average surface brightness of 
$\sim4.4\times10^{-14}$~ergs~cm$^{-2}$~s$^{-1}$~arcsec$^{-2}$. The total integrated \ha\, flux for UM~448 is 
9.24$\times$10$^{-13}$~ergs~cm$^{-2}$~s$^{-1}$, which is in good 
agreement with the value of $\sim11.5\times10^{-13}$~ergs~cm$^{-2}$~s$^{-1}$ 
derived by \citet{Dopita:2002} from narrow band \ha\, imaging. Relative
line intensities are listed in Tables~\ref{tab:um448-1}--\ref{tab:um448-3} for summed spectra over each of the three designated regions, for each separate velocity component.

The \ha\, velocity map (Fig.~\ref{fig:um448_ha}b) displays a rather similar radial velocity structure for both C1 and C2 components. Both show a smooth velocity distribution 
between $+$50 and $-$180~\kms, which suggests solid body rotation within the 
galaxy. These maps can be
used to derive an axis of rotation for the gaseous component of UM448 along
position angle, PA $\sim36^\circ$
(shown as a dashed line in Fig.~\ref{fig:um448_ha}b). Interestingly, the alignment of this axis also divides the blue eastern `tail' from the red, western part of the galaxy (Fig.\,~1).  The radial velocity structure of
component C3 is less straightforward than
its counterparts. In the two regions near the center of UM~448, the
velocities range from $-$100 to $+$100~\kms, whilst in the extreme west of the galaxy they are consistently negative down to $-$300~\kms.  The velocity
structure there could be the signature of a discrete blue-shifted outflow,
which is perhaps related to the detached knots of emission present in the
SuSI2 images $\sim10''$ from the southwest corner of the Argus aperture (Fig.\,~1) -- i.e. we are detecting fast-moving gas, kinematically decoupled from the main body of UM448.

The maps of the \ha\, FWHM  show a slight spatial
correlation with the flux distribution (Fig.~\ref{fig:um448_ha}c).  The eastern regions displays decreased FWHMs for both the C1 and C2 line components, in comparison to the upper arm structure: the average FWHM for the lower arm is
$\sim$40--60~\kms\, for C1 and 120--160~\kms\, for C2, which
increases to $\sim$90--100~\kms\, and 200--240~\kms, respectively,
in the upper arm. Interestingly, the region of low FWHM is co-spatial with the main peak in flux within the eastern arm. If the overall appearance of UM448 is indicative of a close encounter or merger of two systems, then it would make sense for the gas turbulence, and hence the FWHM, to increase as one approaches the transition region between the two colliding components (the blue/red interface in Fig.\,~1 right). On the other hand, component C3 shows a typical FWHM of 70~\kms suggesting an association with more orderly gas volumes.

\begin{figure*}
\includegraphics[scale=0.55]{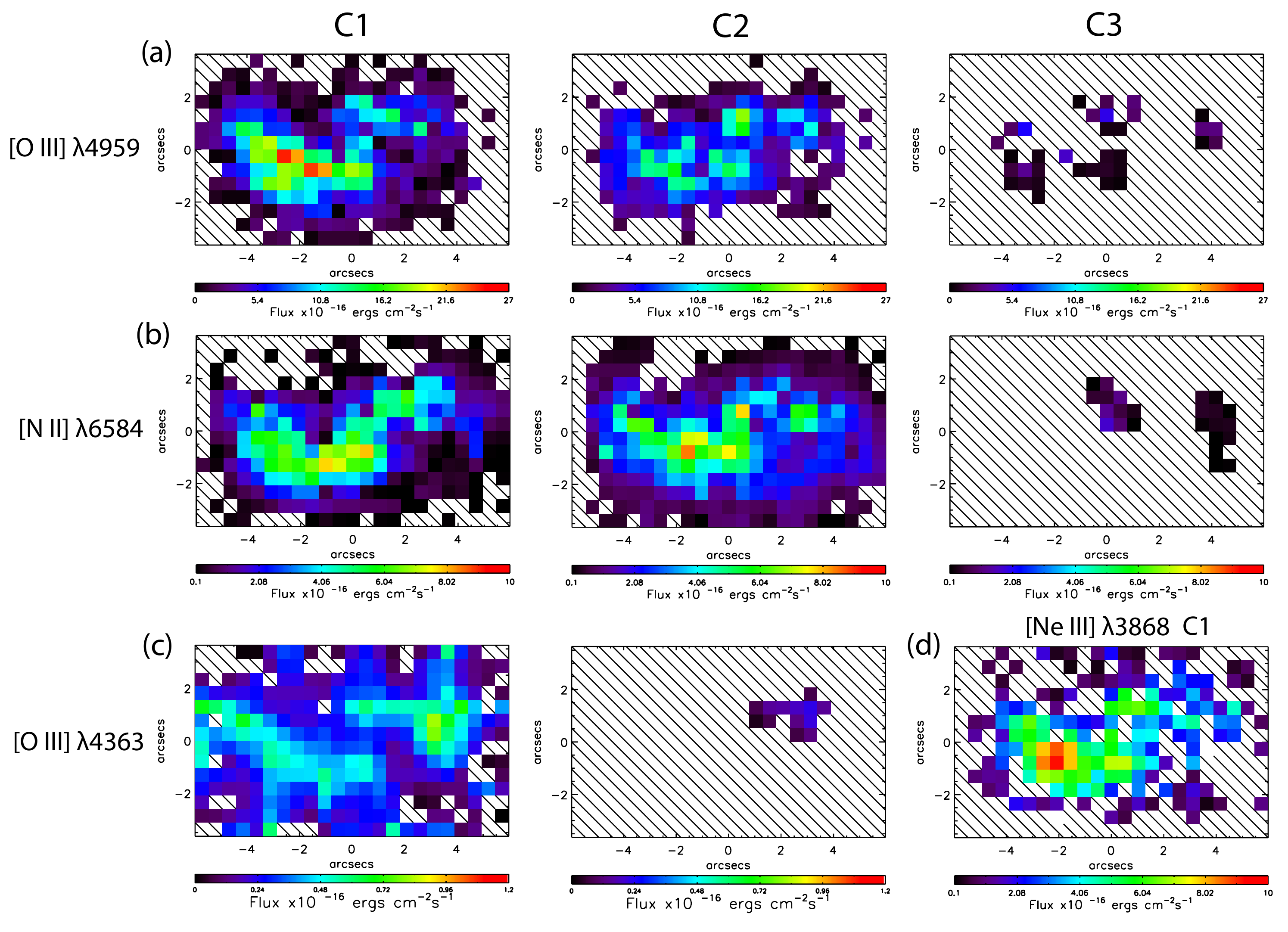}
\caption{Dereddened intensity maps of UM~448 in \foiii~$\lambda$4959, 
\fnii~$\lambda$6584, \foiii~$\lambda$4363 and  \fneiii~$\lambda$3868 for the separate
velocity components C1, C2 and C3. \foiii~$\lambda$4363 is shown after being re-binned by
1.5$\times$1.5 spaxels and re-mapped onto the original grid (see the text for details).
Only spaxels with S/N$\geq$10 are shown for \foiii~$\lambda$4959 and \fnii~$\lambda$6584,
$\geq$5 for  \fneiii~$\lambda$3868 and $\geq$1 for \foiii~$\lambda$4363.  North is up and
east is to the left.}
\label{fig:um448_fmaps}
\end{figure*}

\subsection{Forbidden-line maps}
Emission line maps of UM~448 in the light of the \foiii\, 
\fnii\, and \fneiii\, lines for the various velocity components are shown in Fig.~\ref{fig:um448_fmaps}. The morphology of
\foiii~$\lambda$4959 is similar to that of \ha, with the surface
brightness of the lower arm being higher by a factor of $\sim$2 than that of the upper
arm (for component C1). In contrast, \fnii\, displays a rather constant surface
brightness throughout either arm, with a peak in emission
within our designated region 2. For both \foiii\, and \fnii\, line
component C2 displays a similar morphology to the C1 counterparts, as was
the case with \ha. Component C3 of \foiii~$\lambda$4959 has a wider spatial extent
than its counterpart in \ha.  For \fnii\, component C3 is seen to
overlap roughly with the central and eastern regions observed for \ha\--C3.  In the central regions C3 shows symmetry approximately about the rotation axis defined in Fig 3b; this may a signature of an outflow, or it simply reflects the highly disturbed kinematics in the ``collision zone" of the merging galactic components at the Region1/Region2 aperture interface.
\foiii~$\lambda$4363 has also been detected in velocity components C1 and C2 
whose peak surface brightness is found within region 3 (see Fig.~\ref{fig:um448_SFR3_O3c} for an example fit to the summed spectra across this region), whereas $\lambda$4959  peaks in
region 1. This is mirrored in the high electron temperature of region 3 versus region 1
(see below).
The \fneiii~$\lambda$3868 emission also peaks within region 1, with
more diffuse emission extending throughout regions 2 and 3.

\begin{figure*}
\includegraphics[scale=0.8]{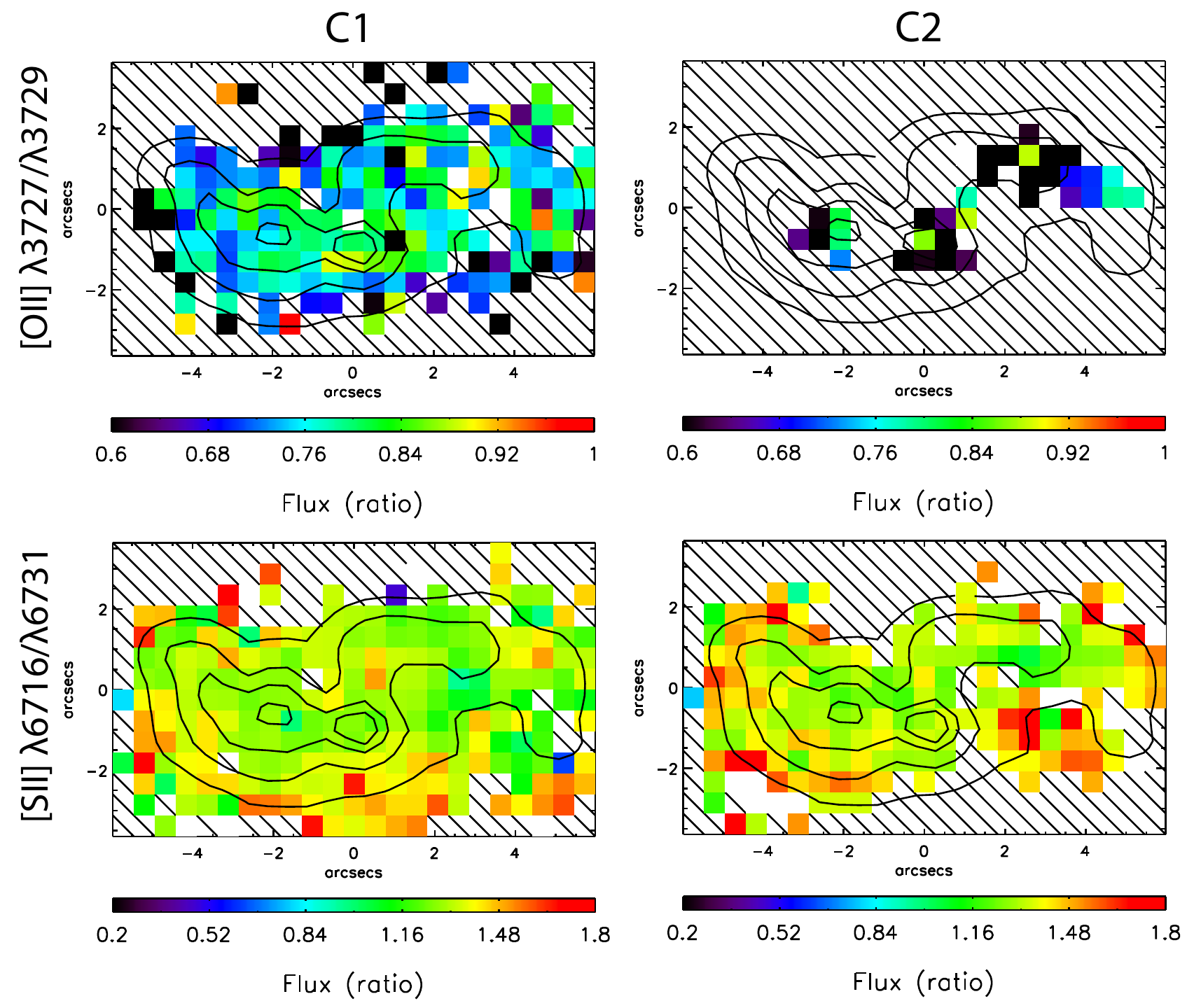}
\caption{Dereddened emission line maps of UM~448 in doublet ratioes 
\foii~$\lambda$3727/$\lambda$3729 (top-panel) and \foii~$\lambda$6716/$\lambda$6731(bottom-panel), in the separate C1(narrow) and C2 (broad) velocity components, with integrated \ha\, emission line flux overlaid in contours. North is up and east is to the left.}
\label{fig:um448_O2S2}
\end{figure*}

\begin{figure}
\includegraphics[scale=0.5]{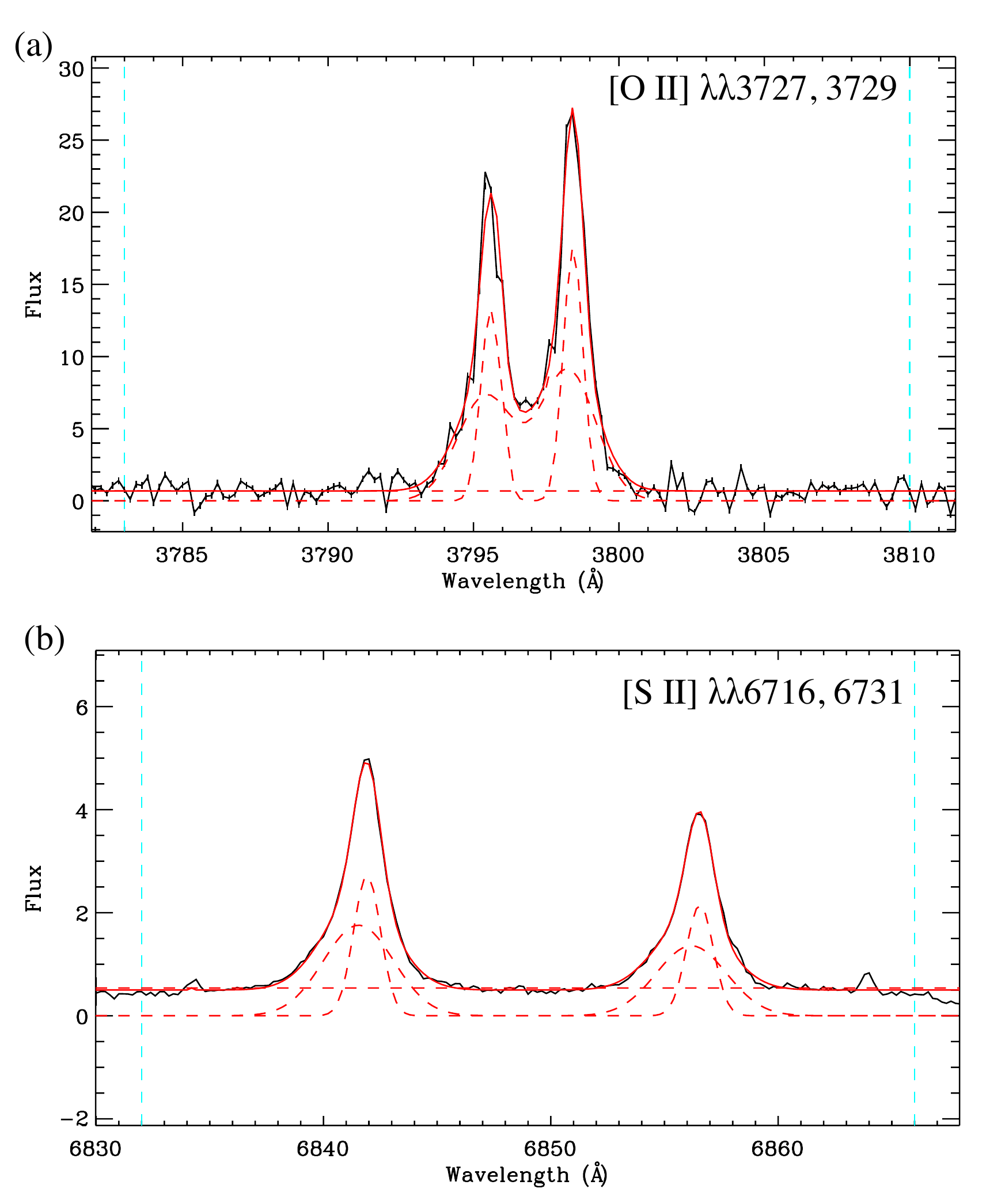}
\caption{Example single-spaxel spectra and fits to the \foii~$\lambda\lambda$3727,3729 and 
\fsii~$\lambda\lambda$6716,6731 lines.  Flux is in units of 
$\times10^{-16}$~ergs~cm$^{-2}$~s$^{-1}$~arcsec$^{-2}$~\AA$^{-1}$ and 
wavelengths are in the rest-frame.}
\label{fig:um448_doublets}
\end{figure}

UM~448 is shown in the light of the \foii\, and \fsii\, doublet lines in 
Fig.~\ref{fig:um448_O2S2}. Due to
their relatively low S/N ratio (compared to the \hi\ Balmer lines and 
\foiii~$\lambda$4959), both doublets could be decomposed to only the C1 
and C2 velocity components (Fig.~\ref{fig:um448_doublets}).  The \foii\, 
doublet, with a rest-wavelength separation of 2.783~\AA\ was particularly 
hard to decompose into a narrow$+$broad component since the large FWHM of
the component C2 would effectively merge the doublet into a single line.  To
alleviate this, each spaxel with a broad component \foii\, line detection
was examined and only selected when there was sufficient S/N ratio to allow for
a de-blended fit (see Fig.~\ref{fig:um448_doublets}a). The larger
wavelength separation of the \fsii\, doublet allowed the broad component 
(C2) to be also fitted in relatively low S/N ratio spaxels (e.g.
Fig.~\ref{fig:um448_doublets}b), and was thus mapped more extensively than \foii\
(compare the lower panels of Fig.~\ref{fig:um448_O2S2}a, b). 
The \foii~$\lambda$3727+$\lambda$3729 emission has a rather disjointed morphology
compared to the integrated \ha\ surface brightness. The ratio of the
\foii\ doublet components commonly used as a density diagnostic, (Fig.~\ref{fig:um448_O2S2},
left panels) displays negligible structure in
relation to the overlaid \ha. In contrast to \foii, the
\fsii~$\lambda$6716+$\lambda$6731 doublet has a morphology that is more akin to \ha\, with three
distinct peaks in flux.  The 
$\lambda$6716/$\lambda$6731 ratio (Fig.~\ref{fig:um448_O2S2}, right panels), 
which is also used as a density diagnostic, decreases in regions that 
align with the peaks in \ha\, emission, and this is indicative of increased electron
density there.

\section{Electron Temperature and Density}
The \foiii\ ($\lambda$5007 $+$ $\lambda$4959)/$\lambda$4363
intensity ratios were used to determine electron temperatures. Two 
emission line doublets are available to determine the electron density, 
\eld; namely the \foii\, $\lambda$3726/$\lambda$3729
and \fsii\, $\lambda$6716/$\lambda$6731 ratios. \elt\ and \eld\ values were computed using
\textsc{iraf's} {\tt ZONES} task in the {\tt NEBULA} package. This makes use of technique that derives both the temperature
and the density by making simultaneous use of temperature- and density-sensitive line
ratios. Atomic transition probabilities for \opp\, and \op\, were taken from \citet{Wiese:1996} whilst collision strengths were taken from \citet{Lennon:1994} and \citet{McLaughlin:1993}, respectively .  For \fsii\, we used the transitional probabilities of \citet{Keenan:1993} and the collisional strengths of \citet{Ramsbottom:1996}.

Where possible \elt\, and \eld\, values were derived for
each velocity component within each of the aperture regions defined in
Fig.~\ref{fig:um448_SFR} and are listed in Table~\ref{tab:um448_ab}. The mean electron
densities from the \fsii\, and \foii\, ratios agree to within $\sim$20\%; the \foii\
density map is however significantly more noisy. The \fsii\
densities were therefore adopted for the analysis. \

\begin{figure*}
\includegraphics[scale=1.0]{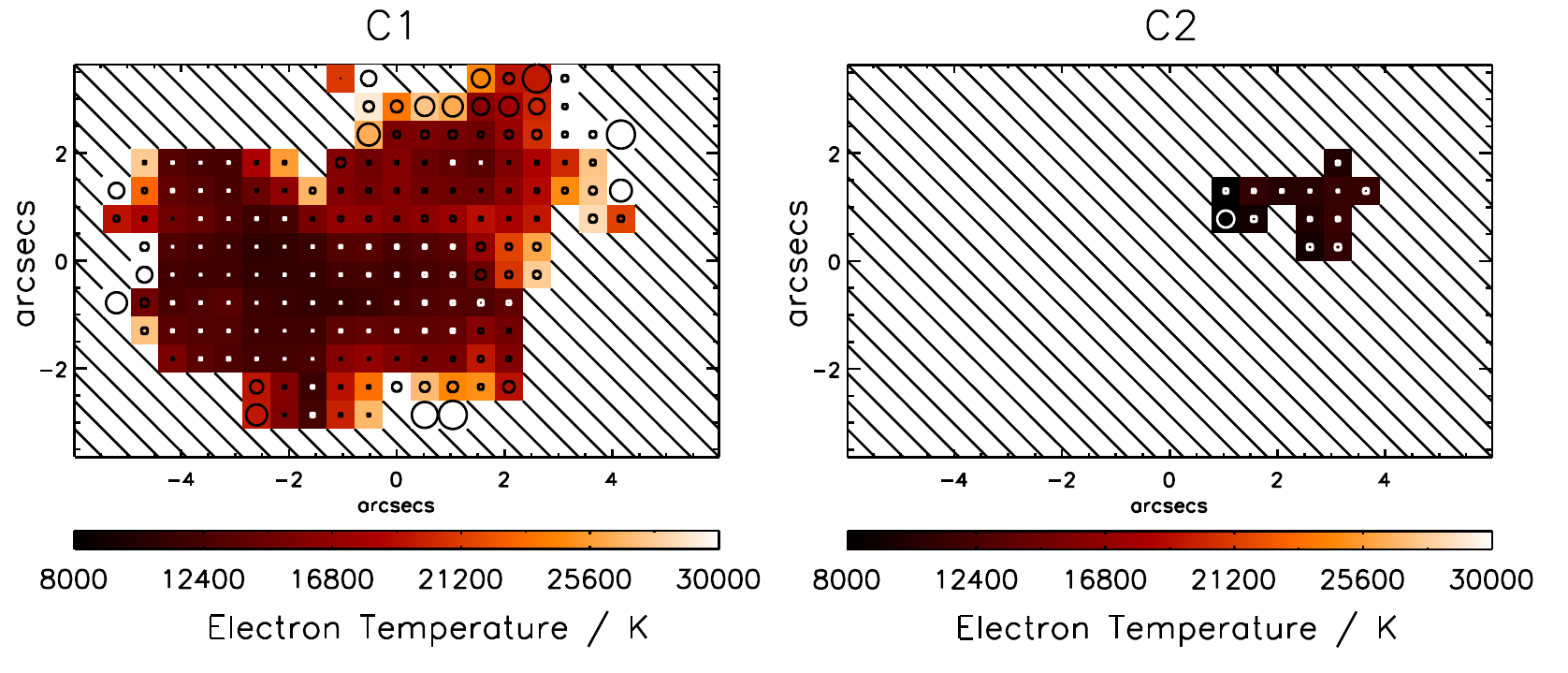}
\caption{\foiii\ electron temperature maps
of UM~448 for C1 (narrow) and C2 (broad) velocity components.  Overlaid circles represent the size of the \elt\, uncertainty within each spaxel (percentage errors are scaled by the 0.52\arcsec\, diameter of a spaxel). North is up and east is to the left.}
\label{fig:um448_tene}
\end{figure*}

\elt\ and \eld\ maps computed for line components C1 and C2 are shown
in Fig.~\ref{fig:um448_tene}, and mean values for each of the aperture regions 1--3 of
Fig.~\ref{fig:um448_SFR} are listed in
Table~\ref{tab:um448_ab}.
Fig.~\ref{fig:um448_tene} (left) shows evidence for a small amount of \elt\, variation
throughout UM~448.  Aperture regions 1 and 2 show temperatures of $\sim$
12,000--13,000\,K. Region~3 shows temperatures of 15,300~K for the narrow line component
C1 and 10,200\,K for the broad line component C2. \eld\, variations are also apparent and
denser areas are correlated with peaks in \ha\ emission.  Regions 1 and 2 have an
average \eld$\sim$120--150~cm$^{-3}$, whereas region 3 is only marginally denser
with 200~cm$^{-3}$.

\section{Chemical Abundances}
\label{sec:abundances}

\begin{figure*}
\includegraphics[scale=0.75]{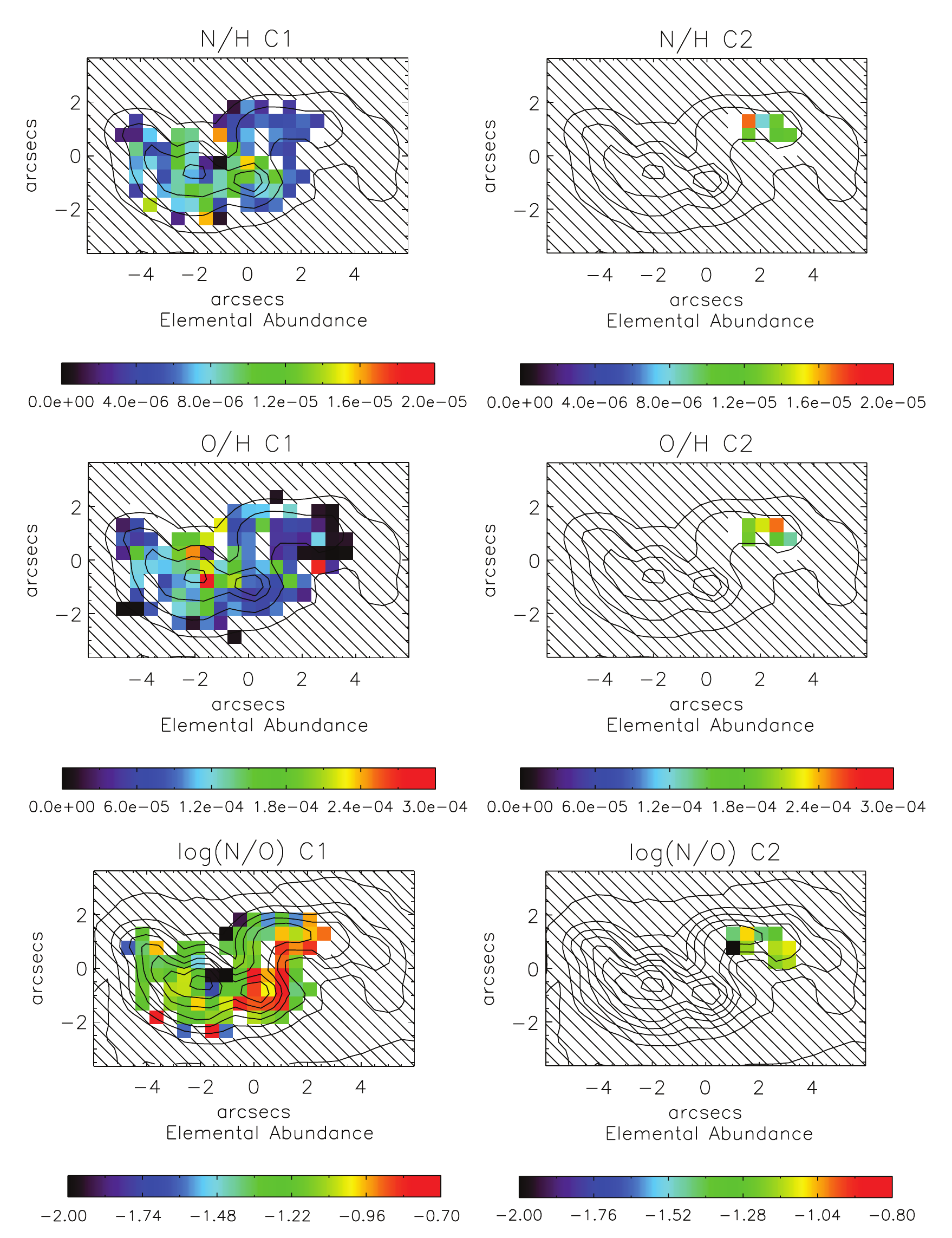}
\caption{UM~448 nitrogen, oxygen and log(N/O) abundance maps for velocity components
C1 and C2, with integrated light \ha\, emission line contours overlaid in dashed line. North is
up and east is to the left.}
\label{fig:um448_ONabund}
\end{figure*}

Ionic abundance maps relative to H$^+$ were created for the \np, \op,
\opp, \nepp\, and \Sp\, ions, using the $\lambda\lambda$6584, 3727$+$3729, 
4959, 3868 and 6717$+$6731 lines respectively. Examples of the flux maps 
used in the derivation of these ionic abundance maps can be seen in 
Figure~\ref{fig:um448_fmaps}. Abundances were calculated using the \textsc{zones} task in
\textsc{IRAF}, using the respective \elt\, and \eld\, maps described 
above, with each FLAMES spaxel treated as a distinct `nebular zone' with 
its own set of physical conditions.  These results were compared with those obtained
using the multilevel ionic equilibrium program \textsc{equib} (originally written by
I.~D.~Howarth and S.~Adams at Univ. College London), which comprises a different set of
atomic data tables. Typical differences in the abundance ratios were found to be 5\% or
less.  To check the validity of using \elt (\foiii) for the singularly ionized elements (\np, \Sp, and \op), we also computed abundance maps useing a \elt (\foii) map (created using the relationship with \elt (\foiii) described in \citet{Izotov:2006}). Discrepancies between the two temperatures were found to be low ($\sim$100--400~K) and typical differences in ionic abundance were $\sim$2--7\%.  These errors are lower than those caused by uncertainties in \elt (\foiii) alone and we therefore do not include it as a source of error.

Ionic nitrogen, neon and sulphur abundances were converted into N/H, Ne/H 
and S/H abundances using ionisation correction factors (ICFs) from 
\citet{Barlow:1994}.The O/H abundance was
obtained by adding the \opp /\hp\, and \op /\hp\, abundance maps (i.e. assuming that
the O$^{3+}$/\hp\ ratio is negligible). Since
the \fsiii~$\lambda$6312 line was not within the wavelength range of our 
data, the \Spp/\hp\, abundance was estimated using the empirical relationship between the
S$^{2+}$ and S$^+$ ionic fractions from the corrected equation
A38 of \citet{Barlow:1994}, namely S$^{2+}$/S$^+$ = $4.677 \times (O^{2+}/O^+)^{0.433}$.
The S/H results should therefore be considered with caution, but they are not at variance with those obtained using the ICF prescriptions of Izotov et al. (2006). 
O/H, N/H and log(N/O) abundance maps are shown in 
Fig.~\ref{fig:um448_ONabund} and are
described below.  Regional abundances derived directly from taking 
error-weighted averages over the abundance maps are listed in 
Table~\ref{tab:um448_ab}. 

\subsection{Maps}
The elemental oxygen abundance map shown in the middle panel of 
Figure~\ref{fig:um448_ONabund} displays a single peak whose location 
correlates with the peak in \ha\, relating to region 1.  Minimal 
abundance variations are seen across UM~448, with a slightly 
decreasing oxygen abundance as one moves from regions 1--3, with an overall average abundance for C1 of $12+log(O/H)=8.15\pm0.10$.  Adopting a solar 
oxygen abundance of 8.71$\pm$0.1 relative to hydrogen \citep{Scott:2009}, 
this corresponds to $\sim$0.28~\Zsol.  This is comparable to the value obtained by \citet[][IT98 hereafter]{Izotov:1998}, who measured an abundance of $12+log(O/H)=7.99\pm$0.04. The C2 velocity component displays a higher oxygen abundance than the C1-component gas within region~3 by $\sim$0.4~dex, which, judging from the agreement between C1 and C2 \foiii\, 4959\,\AA\ flux (Fig.~\ref{fig:um448_fmaps}) is mostly attributable to the \elt\, being $\sim$5000~K cooler (if the C1 region was instead much denser but of similar oxygen abundance to the gas sampled by the C2 line component one might expect the $\lambda$4959/H$\beta$ C1 ratio to be much smaller something which is not observed -- Table A3).

The upper panel of Fig.~\ref{fig:um448_ONabund} shows the N/H abundance 
across UM~448.  The C1 peak in nitrogen abundance is offset from that of 
oxygen, peaking at the location of region 2 with an average N/H ratio 
of 1.04$\pm0.01 \times10^{-5}$.  For region 1 it is slightly lower at 
9.40$\pm0.02\times10^{-6}$ and for region 3 even lower at 7.00$\pm0.03\times10^{-6}$.  The C2 component gas in region 3 has a nitrogen abundance that is higher than its C1 counterpart by a factor of 2.
As a result, there is a peak in the N/O ratio within UM~448 located at 
region 2, which can be seen in the bottom panel of 
Fig.~\ref{fig:um448_ONabund}.  For C1, regions 1 and 3 agree with an average of log(N/O) 
$\sim$ --1.23$\pm$0.04, whereas region 2 has an average (N/O) ratio of $\sim$ 
--1.03$\pm$0.05.  The map of log(N/O) shows that this peak actually reaches 
$\sim -0.80$ at the precise location of region 2 (as can be seen from 
the \ha\, contours).  The N/O ratio of the (broad) C2 component gas in region 3 ($\sim$ --1.26$\pm$0.13) is in good agreement with the average N/O of C1 gas in that region.  The long-slit study by IT98 measured an N/H ratio of 
$9.5\pm2.0\times10^{-6}$, in agreement with the averages of regions 1 and 2 
derived from the N/H map shown in Fig.~\ref{fig:um448_ONabund}.  The average C1
N/O ratios for region 3 is slightly higher (by $\sim$0.2~dex) 
than those of other galaxies of similar oxygen 
metallicity \citep[see][their figure 11]{Lopez-Sanchez:2010}. They are, 
however, consistent within the uncertainties.  Region 2, however, has a 
higher value than the average value of $\sim-1.4\pm0.2$ expected for 
its metallicity, even when taking the uncertainties into consideration, 
for both the averaged value of region 2 and on a spaxel-by-spaxel 
basis. For this region we do observe the factor of $\sim3$ nitrogen excess 
reported by \citet{Pustilnik:2004}.

The Ne/H and S/H 
abundance distribution throughout UM~448 for the narrow (C1) emission line 
component show only minimal abundance 
variations in relation to the \ha\, emission.  Ne/H values averaged over each region range from 
$\sim$2.06--6.55$\times10^{-5}$, with the highest Ne/H abundance aligning with region 2. The 
Ne/O ratios of region 1 and 3 are typical for those expected by BCGs at similar 
metallicities; the averaged region values lie within the reported 
range of log(Ne/O) $\sim-0.9$ to $-0.7$ \citep{Izotov:2006}.  However, the Ne/O ratio for region 2 is $\sim$0.5~dex higher than the expected Ne/O value. 
Calculations of Ne/H were also made using the Ne$^{2+}$ ICF of \citet{Izotov:2006}, however the Ne/O ratio for region 2 remained $\sim$0.2~dex higher than the expected value.
The S/H distribution 
across UM~448 is relatively constant for the C1 and C2 gas components, with each region displaying 
averages within the range of $\sim4.7$ to $7.4\times10^{-6}$.  The C1 S/O ratio for all regions are 
slightly higher than expected for blue compact galaxies, log(S/O)$\sim -1.9$ to $-1.5$ \citep{Izotov:2006}, and the S/O value C2 component in Region~3 lies at the upper limit of this average. 

\begin{figure}
\includegraphics[scale=0.45]{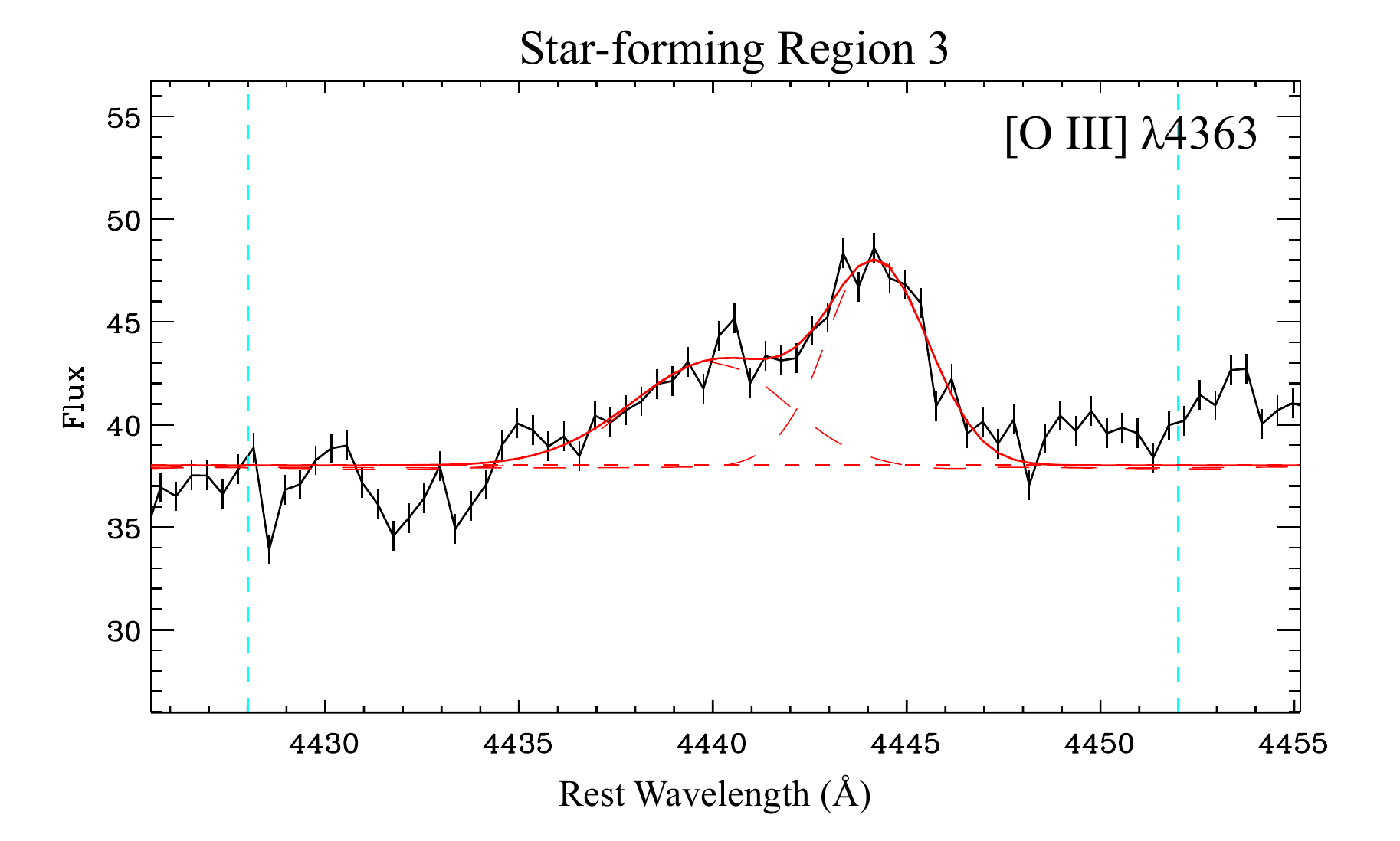}
\caption{\foiii~$\lambda$4363 profile observed in the spectra summed over region 3,
displaying a narrow (C1) plus broad (C2) component Gaussian profile.  Flux, FWHM and
velocity parameters for each component are listed in Table~\ref{tab:um448-3}}
\label{fig:um448_SFR3_O3c}
\end{figure}

\subsection{Comparison of spatially resolved and global spectra abundances}

Also listed in Table~\ref{tab:um448_ab} are the \elt, \eld, ionic and elemental abundances derived from summed spectra over the entire galaxy (using line fluxes listed in Table~\ref{tab:um448_sum}). This particular set of results allows us to assess if global galaxy spectra can reliably represent the physical properties of its ISM. For example, values derived from long-slit observations may contain luminosity-weighted fluxes and consequently erroneous ratios, or large apertures may include a mixture of gas with different ionization conditions and/or metallicities. This question was addressed by \citet{Kobulnicky:1999} to assess whether global emission-line spectra can be depended upon to reliably recover the chemical properties of distant star-forming galaxies. Overall, they find that when using direct-method techniques, global spectra can give an accurate representation of the oxygen abundances derived from individual \hii\, regions, but only after applying a correction of $\Delta$(O/H)$=0.1$ to the global spectra results. In addition to this, \citet{Pilyugin:2012} found that oxygen abundances were systematically underestimated (and N/O consequently over-estimated) when using global SDSS (Sloan Digital Sky Survey) spectra of \hii\, regions with different physical properties.

For UM~448, the temperature derived from the integrated spectrum is in agreement with the flux-weighted mean \elt\, across the three regions ($14010\pm2100$~K), suggesting that the integrated \foiii\, ratio may have been more heavily weighted by areas of stronger \foiii~$\lambda$4363 emission.  The oxygen abundance from the integrated spectrum, 12$+$log(O/H)=8.34$\pm$0.07, is higher than all C1-component values, suggesting that it may be weighted towards the high O/H abundance seen in the C2-component of region 2.  Being $+0.18$~dex higher than the flux-weighted average of the three regions (12+log(O/H)=8.16$\pm$0.18), the globally-derived (i.e. derived from the integrated spectrum) O/H is not fully representative of the emission from the whole galaxy, even after applying the correction suggested by \citet{Kobulnicky:1999}. Nitrogen too, displays a higher abundance when derived from the summed spectrum rather than the regional averages, and results in a (log) nitrogen-to-oxygen ratio of $-$1.15, which is approximately the expected value at this metallicity.  The Ne/H abundance ratio from the integrated spectrum is almost a factor of 2 higher than the flux-weighted average of the three regions ($\sim$2.14$\times10^{-5}$), whilst the log(Ne/O) ratio is in good agreement. The same applies for the S/H abundance, although less severe, where the globally-derived S/H ratio is a factor $\sim$1.5 higher than the flux-weighted average S/H across the three regions, but the log(S/O) value is in agreement between the two methods. 

Comparisons can also be made with the results from a long-slit study by IT98, which utilized summed spectra over a 2\arcsec$\times$7\arcsec\, region `across the brightest region of the galaxy'.  From this spectrum they derive an \elt\, of 12100$\pm$500~K (in line with the C1 component values of our region 1), O, N, Ne and S, abundances that are systematically lower than those derived from our summed spectrum and abundances relative to oxygen that are in relatively good agreement within the uncertainties. However, without knowing the exact positioning of this slit, a truly meaningful comparison cannot be made.

Overall it can be said that whilst abundance ratios of heavy elements can be reasonably well obtained by summed spectra and/or long-slit observations, the ionic and elemental abundances relative to hydrogen are not well reproduced; this may be because the former properties do not have a very severe sensitivity to \elt\ biases, while the latter do. This is in addition to the fact that, in contrast to integral field spectroscopy, global properties derived from light-averaged methods cannot resolve abundance variations and any potential sites of enrichment are being averaged out.  Of course, uncertainties will still exist within IFU observations due to averaging along the line of sight of individual spaxels; this is something that can only be remedied with the development of higher spectral resolution IFUs.

\begin{table*}
\begin{center}
\begin{footnotesize}
\begin{tabular}{|l|c|c|cc|c|}
\hline
                &\multicolumn{1}{c|}{Region~1}      &\multicolumn{1}{c|}{Region~2}      &\multicolumn{2}{c|}{Region~3}& Summed Spectra \\
\hline
        &       C1                                      &       C1      &C1     &C2& All Components \\
(1) & (2) & (3) & (4) &(5) & (6)\\

\hline   
\elt\ & 12220 $\pm$ 60 & 13020 $\pm$ 150 & 15340 $\pm$ 190 & 10230 $\pm$ 210 &                                                                               13,610	$\pm$	560	\\
\eld\ & 100 $\pm$ 25 & 80 $\pm$ 35 & 115 $\pm$ 35 & 95 $\pm$ 30&                                                                                             40	$\pm$	6	\\
 \noalign{\vskip4pt}                                                                                                                                         			  
O$^+/H^+$ $\times10^5$ &         7.24$\pm$        0.27 &         6.50$\pm$        0.53 &         5.23$\pm$        0.53 &         15.49$\pm$      1.55 &      12.25	$\pm$	0.62	\\
O$^{2+}$/H$^+$ $\times10^5$ &         9.00$\pm$        0.31 &         4.56$\pm$        0.31 &         5.24$\pm$        0.43 &         10.91$\pm$   0.33&     9.40	$\pm$	0.48	\\
O/H $\times10^4$ &         1.62$\pm$        0.08 &         1.11$\pm$        0.12 &         1.05$\pm$        0.14 &            2.64$\pm$         0.28 &       2.16	$\pm$	0.16	\\
12+log(O/H) &         8.21$\pm$        0.05 &         8.04$\pm$        0.11 &         8.02$\pm$        0.13  &  8.42$\pm$        0.10&                       8.34	$\pm$	0.07	\\
Z/\Zsol &         0.32$\pm$        0.02 &         0.22$\pm$        0.02 &         0.20$\pm$        0.03 &                     0.51$\pm$         0.05  &      0.42	$\pm$	0.03	\\
 \noalign{\vskip4pt}                                                                                                                                         			  
 N$^+/H^+$ $\times10^6$ &         4.19$\pm$        0.14 &         6.08$\pm$        0.28 &         3.49$\pm$        0.21 &          8.45$\pm$      0.85 &     8.74	$\pm$	0.76	\\
ICF(N)			 &2.24					&1.70				&2.00				&1.70		&            1.77		\\
N/H $\times10^6$ &         9.40$\pm$        0.66 &        10.35$\pm$        1.46 &         7.00$\pm$        1.24 &             14.40$\pm$       2.53 &       15.44	$\pm$	1.91	\\
log(N/O) &        -1.24$\pm$        0.05 &        -1.03$\pm$        0.09 &        -1.18$\pm$        0.12 	&$-$1.26$\pm$  0.14 &                        -1.15	$\pm$	0.10	\\
 \noalign{\vskip4pt}                                                                                                                                         			  
 Ne$^{2+}/H^+$ $\times10^5$ &         2.10$\pm$        0.12 &         2.70$\pm$        0.5 &         1.03$\pm$        0.13 & --&                             3.16	$\pm$	0.63	\\
ICF(Ne) & 1.80 & 2.42 & 2.00 & --&                                                                                                                           2.30		\\
Ne/H $\times10^5$ &         3.78$\pm$        0.31 &         6.55$\pm$        0.1.46 &         2.06$\pm$        0.41 & --&                                    7.27	$\pm$	1.59	\\
log(Ne/O) &        -0.63$\pm$        0.10 &        -0.23$\pm$        0.25 &        -0.71$\pm$        0.24 & --&                                              -0.47	$\pm$	0.20	\\
 \noalign{\vskip4pt}                                                                                                                                         			  
S$^+$/H$^+$ $\times10^6$ &         1.11$\pm$        0.04 &         0.92$\pm$        0.05 &         1.06$\pm$        0.06 &         1.43$\pm$      0.09 &     1.87	$\pm$	0.17	\\
ICF(S)			   &1.06				&1.02				&1.04				&1.02	&                    1.03		\\
S/H $\times10^6$ &         7.23$\pm$        0.45 &         4.75$\pm$        0.63 &         6.30$\pm$        1.05  &   7.35$\pm$        1.06 &                9.93	$\pm$	0.92	\\ 
log(S/O) &    -1.35$\pm$        0.08 &        -1.37$\pm$        0.17 &        -1.22$\pm$        0.21  &            -1.56$\pm$        0.18 &                  -1.34	$\pm$	0.12	\\ 
 \noalign{\vskip4pt}                                                                                                                                         
                                                                                                                                                             
\hline
\end{tabular}
\caption{Ionic and elemental abundances for UM~448, with the corresponding \elt\, and \eld\, used in their derivation.  Columns 2--5 refer to abundances derived from maps, weight-averaged over regions defined in Fig.~\ref{fig:um448_SFR}.  Column 6 refers to \elt, \eld\, and abundances derived from spectra integrated over UM~448 (corresponding fluxes are listed in Table~\ref{tab:um448_sum}).  All ionisation correction factors (ICFs) are taken from \citet{Barlow:1994}.  S/H ratios were calculated using the ICF(S) values listed along with a predicted value of \Spp\, (see text for details).}
\label{tab:um448_ab}
\end{footnotesize}
\end{center}
\end{table*}



\section{Stellar Populations}

\subsection{Age of the Stellar Population}
\begin{figure}
\includegraphics[scale=.45]{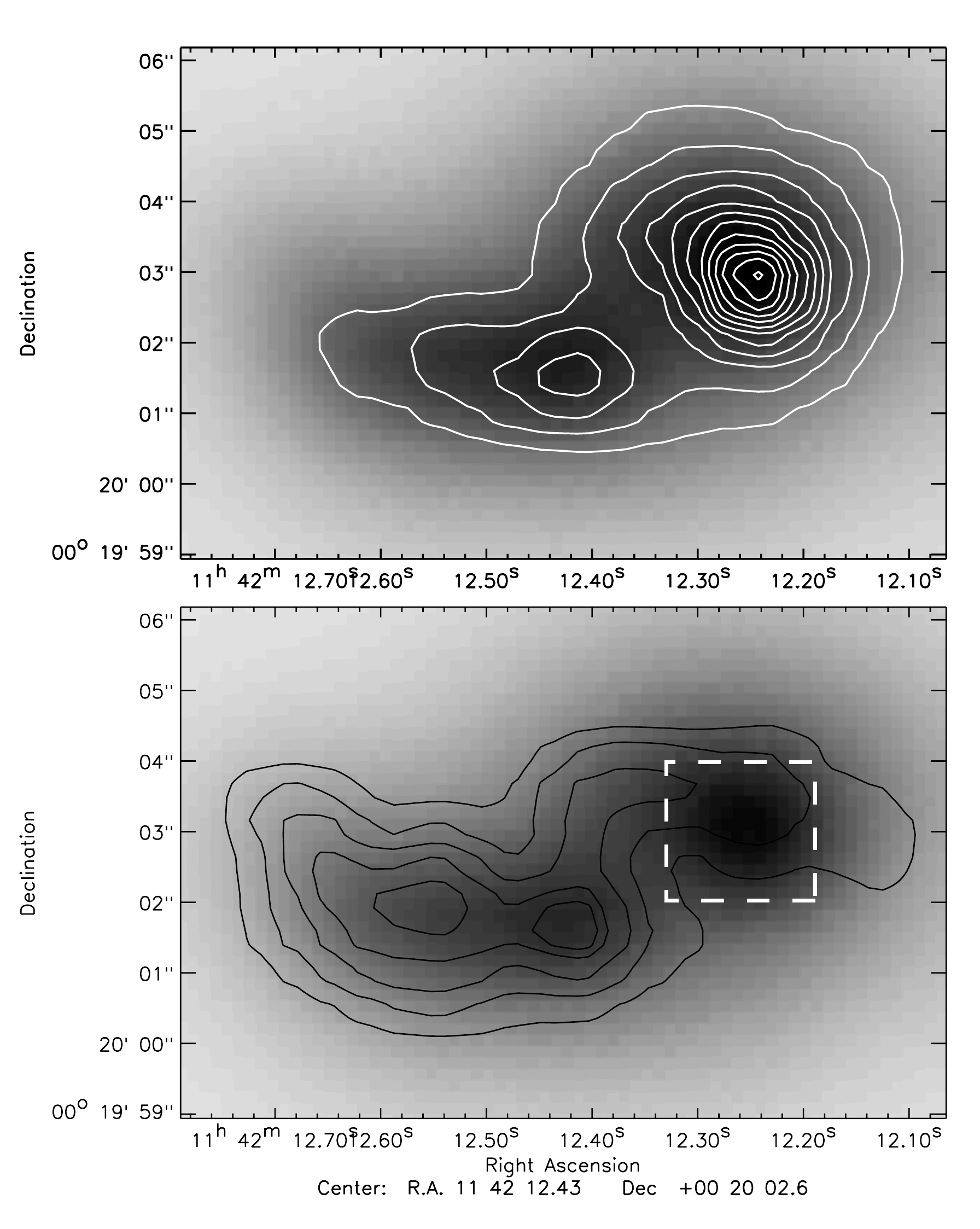}
\caption{NTT-SuSI2 combined $R, V$ and $B$-band images of UM~448 with 
contours showing the continuum near H$\alpha$ 
(\textit{top panel}) and integreated H$\alpha$ emission (\textit{bottom panel}; 
see Fig.~\ref{fig:um448_ha} for maps of the separated emission line 
components).  North is up and east is to the left.  \ha\, contours are 
shown for the range 35.0 to 128.6 in steps of 22.0, in flux units 
of 10$^{-16}$ ergs s$^{-1}$ cm$^{-2}$ arcsec$^{-2}$.  The continuum 
peak region is represented by the white-dashed box; spectra were summed 
over this region to investigate stellar absorption features and averages 
in $EW$(\hb).}
\label{fig:um448_cont}
\end{figure}

Figure~\ref{fig:um448_cont} shows a combined NTT-SuSI2 $R, V$ and $B$-band 
image that shows the stellar continuum within UM~448.  
Overlaid on the top panel are the contours of the continuum emission in the \ha\ spectral vicinity ($\sim$6~\AA\, windows either side of the line) that are in 
good agreement with the morphology of the broad-band images. In the lower 
panel we show the same broad-band images with the \ha\, integrated-light 
emission line contours. Significant differences between the morphology of 
the ionised gas emission and that of the stellar continuum are clearly 
apparent, especially in the top, western `arm'. A strong continuum peak 
can be seen, which is not mirrored by the \ha\, emission, suggesting
that a strong stellar component exists within this region.  From the broad-band 
colour image shown in Figure~\ref{fig:um448_susi} it can be seen that this 
peak in continuum is red in colour, signifying an older stellar 
population than that seen in the bluer, eastern `arm'. Such a distinct, 
two component stellar structure, in conjunction with the difference in 
radial velocity between them (see Fig.~\ref{fig:um448_ha}b), supports 
the interpretation that UM~448 is a merging system comprised of two 
galaxies, with very different ages. Spectra summed over this 
continuum peak region (represented by the white-dashed box overlaid in 
Fig.~\ref{fig:um448_cont}-lower panel) do not reveal any stellar absorption 
features around the \ha, \hb, \hg\, or \hd\, lines, however Ca~\textsc{II} H, K ($\lambda\lambda3970, 3934$) interstellar absorption features were detected.

\begin{figure}
\includegraphics[scale=.50,angle=90]{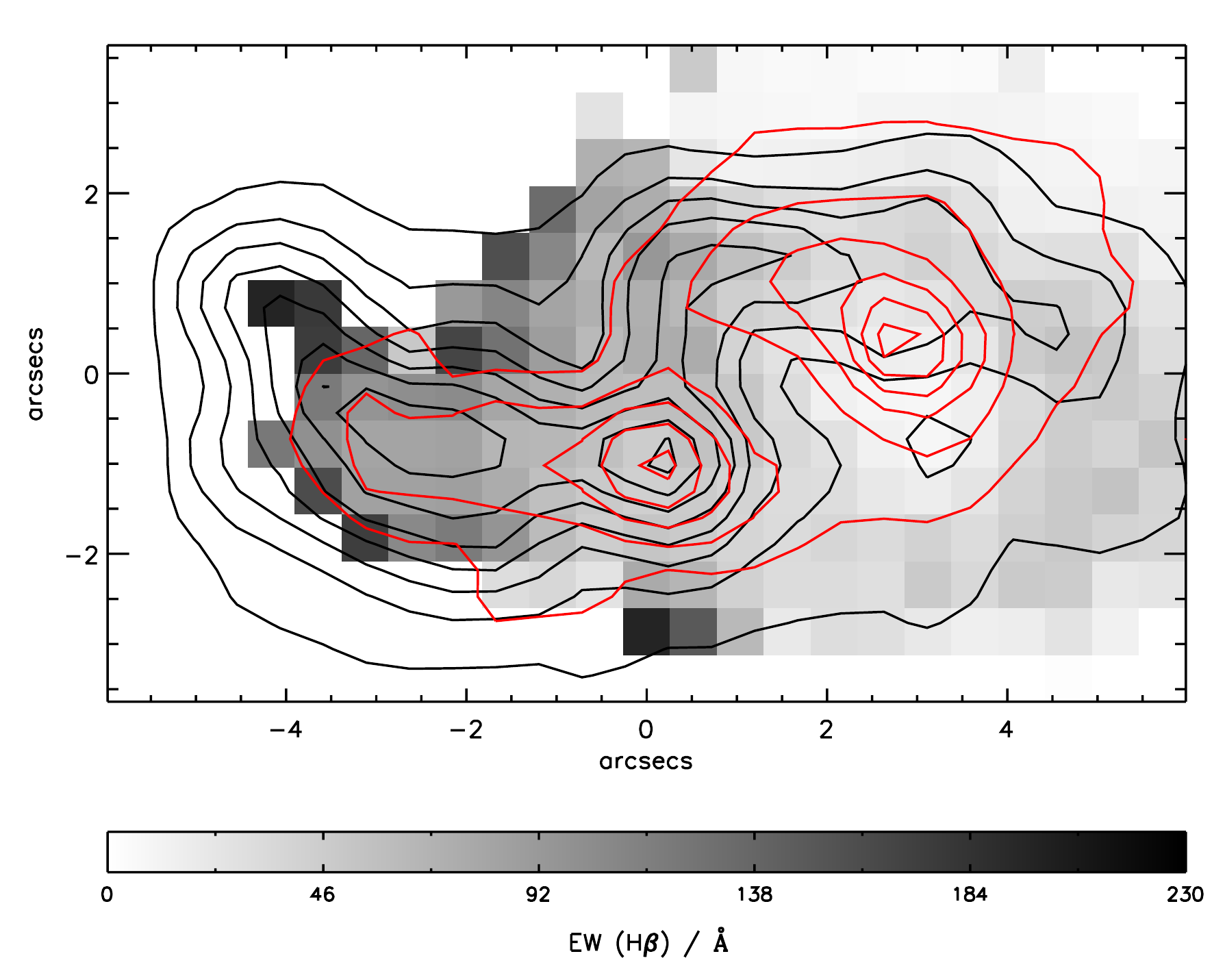}
\caption{Map of the equivalent width of \hb\, accross UM~448.  Overlaid red lines are 
contours of \hb-region continuum; black lines are \hb\, flux contours from
Fig.~\ref{fig:um448_SFR}.}
\label{fig:um448_EW}
\end{figure}

The age of a young population in a galaxy can be estimated using hydrogen 
recombination lines, since they 
provide an estimate of the ionizing flux present, 
assuming a radiation-bounded HII region \citep{Schaerer:1998}.  In 
particular, the equivalent-width (EW) of \hb\, can be used as an age 
indicator of the ionizing stellar population at a given metallicity.  A 
map of $EW$(\hb) for UM~448 can be seen in Figure~\ref{fig:um448_EW}.  
Whilst there are no clear peaks in $EW$(\hb), the main region of 
decreased $EW$ correlates with the largest peak in continuum emission 
(red contours) and that corresponds to region 3.

\begin{figure}
\includegraphics[scale=.55,angle=90]{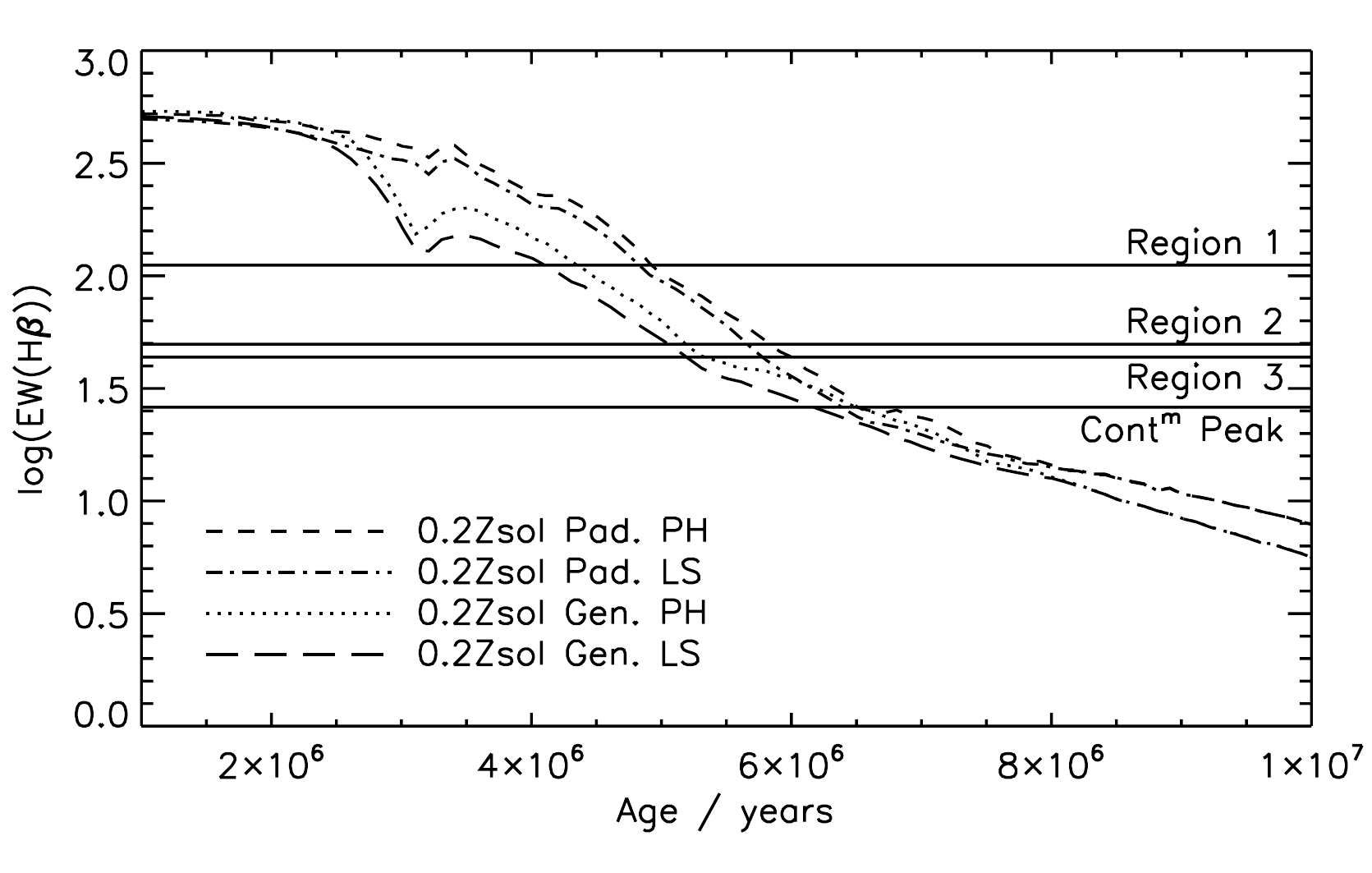}
\caption{$EW$(\hb) as a function of age, as predicted by the \textsc{starburst99} code 
for metallicities of 0.2\Zsol\, using a combination of Geneva or Padova stellar
evolutionary tracks and Lejeune-Schmutz (LS) or Pauldrach-Hillier (PH) model atmospheres. 
The observed average $EW$(\hb)'s for each star forming region in UM~448, along with the
peak in continuum highlighted in Fig.~\ref{fig:um448_cont} (lower panel), are overlaid
(solid line).}
\label{fig:um448_ages}
\end{figure}

Following the method outlined by \citet{James:2010}, we can use this map, in conjunction 
with the metallicity map described in Section
\ref{sec:abundances}, to estimate the age of the most recent star forming 
episodes throughout UM~448, by comparing the regional observed average
$EW$(H$\beta$) values with those predicted by the spectral synthesis code
STARBURST99 \citep{Leitherer:1999}. For the models we chose a metallicity 0.2\,\Zsol,
the closest representative to the weighted average metallicity of UM~448 ($\sim$0.25\,\Zsol, derived from regional averages), together with assumptions of an
instantaneous burst with a Salpeter initial mass function (IMF), a total 
starburst mass
of 1$\times10^6$~\Msol, a
100~\Msol\ stellar upper mass limit (which approximates the classical
\citet{Salpeter:1955} IMF). Models were run for Geneva tracks with ``high''
mass loss rates and Padova tracks with
thermally pulsing AGB stars included. For each of these evolutionary tracks,
two types of model atmosphere were used; firstly the Pauldrach-Hillier (PH) and secondly Lejeune-Schmutz (LS). The latter was
chosen because it incorporates stars with strong winds, which would be
representative of the WR population within the galaxy (if present).  The
stellar ages predicted by each model and the observed average $EW$(H$\beta$)
within each peak emission region in UM~448 were derived from
Figure~\ref{fig:um448_ages} and are listed in Table~\ref{tab:EWs}. 

The difference between the ages predicted by the Geneva and Padova stellar evolutionary
tracks is relatively small, with the Geneva tracks predicting lower ages by up to
23\%. The difference in ages predicted by the PH and LS atmospheres are smaller
still, with LS predicting lower ages by $\sim$5\%.  We cannot comment on which
atmosphere model is more appropriate given that the existence of WR stars
within UM~448 is not confirmed by the current study, hence we adopt 
average stellar ages from the four model combinations. The current (instantaneous) star
formation rates (SFRs) based on the H$\alpha$ luminosities were calculated
following \citet{Kennicutt:1998} and are given in Table~\ref{tab:EWs}. SFRs
corrected for the sub-solar metallicities are also given,
derived following the methods outlined by \citet{Bicker:2005}.

Regions 1--3 contain ionizing stellar populations with average ages of 
4.6, 5.5 and 6.0~Myr, respectively, and the region of peaked continuum 
(Fig.~\ref{fig:um448_cont}) has an average age of 8.2~Myrs 
(Table~\ref{tab:EWs}).  since Region~1 shows the highest ionizing flux, we 
would expect it to contain the youngest stellar population.  It is notable 
that the ages of the stellar population increase through Regions~1--3 as 
one moves away from the blue continuum peak (Fig.~\ref{fig:um448_susi}).  
However, the SFR does not show a smooth decrease in the same direction, 
decreasing slightly for region 2 before increasing again for region 
3.  Interestingly, and not surprisingly, the main red-colored
stellar continuum peak (Figs.~\ref{fig:um448_susi} and 
\ref{fig:um448_cont}) 
displays an increase in age of almost 4~Myrs from the blue continuum peak
(region 1) and a much decreased SFR compared to the three main 
regions of star-formation.  Using narrow-band \ha\, imaging, 
\citet{Dopita:2002} derived a total SFR of 6.6~\Msol/yr, which is 
in good agreement with that derived from our total \ha\ 
flux, 5.0~\Msol/yr.  Accounting for its $\sim$0.2\,\Zsol metallicity, this 
translates to 2.8~\Msol/yr. The evidence suggests that UM~448 is 
made 
up of two separate bodies undergoing a merger; the first is sampled by our aperture 
regions 1 and 2, where relatively recent bursts of 
star-formation have occurred. The second body with which it is merging may be an older 
 galaxy (represented by the red continuum peak) sampled by aperture 
region 3, within 
which the collision could have triggered star-formation, hence an 
increased SFR but overall age that is representative of an aggregate old$+$young population.  


\begin{table*}
\begin{center}
\begin{small}
\caption{Age of the latest star formation episode and current star formation rates.}
\begin{tabular}{lcccc}
\hline
 & \multicolumn{4}{c}{Starburst ages (Myr)}\\
 \cline{2-5}
Model &  Reg.~1 & Reg.~2 & Reg.~3 & Cont$^m$ Peak \\
Padova-AGB PH &        5.06$\pm$       0.24 &        5.86$\pm$       0.13 &        6.36$\pm$       0.48 &        8.61$\pm$       0.74\\
Padova-AGB LH &        4.96$\pm$       0.24 &        5.66$\pm$       0.13 &        6.01$\pm$       0.33 &        8.51$\pm$       0.79\\
Geneva-High PH &        4.41$\pm$       0.27 &        5.31$\pm$       0.16 &        5.91$\pm$       0.56 &        7.91$\pm$       0.74\\
Geneva-High LS &        3.86$\pm$       0.48 &        5.11$\pm$       0.16 &        5.61$\pm$       0.50 &        7.76$\pm$       0.82\\

\\
& \multicolumn{4}{c}{Star formation rates (M$_\odot$ yr$^{-1}$)}\\
\cline{2-5}

SFR(H$\alpha$)$^a$              & 1.49$\pm$0.03 & 0.75$\pm$0.02 & 0.90$\pm$0.02 & 0.34$\pm$0.01 \\
SFR(H$\alpha$)$^b$              &  0.82$\pm$0.02 & 0.41$\pm$0.01 & 0.50$\pm$0.01 & 0.18$\pm$0.01\\
\hline \label{tab:EWs}
\end{tabular}
\begin{description}
\item[$^a$] Derived using the relationship between SFR and $L$(H$\alpha$) from
\citet{Kennicutt:1998}.

\item[$^b$] Corrected for sub-solar metallicities following \citet{Bicker:2005}.

\end{description}
\end{small}
\end{center}
\end{table*}

\subsection{WR stars and N-enrichment}
\begin{figure}
\includegraphics[scale=.5]{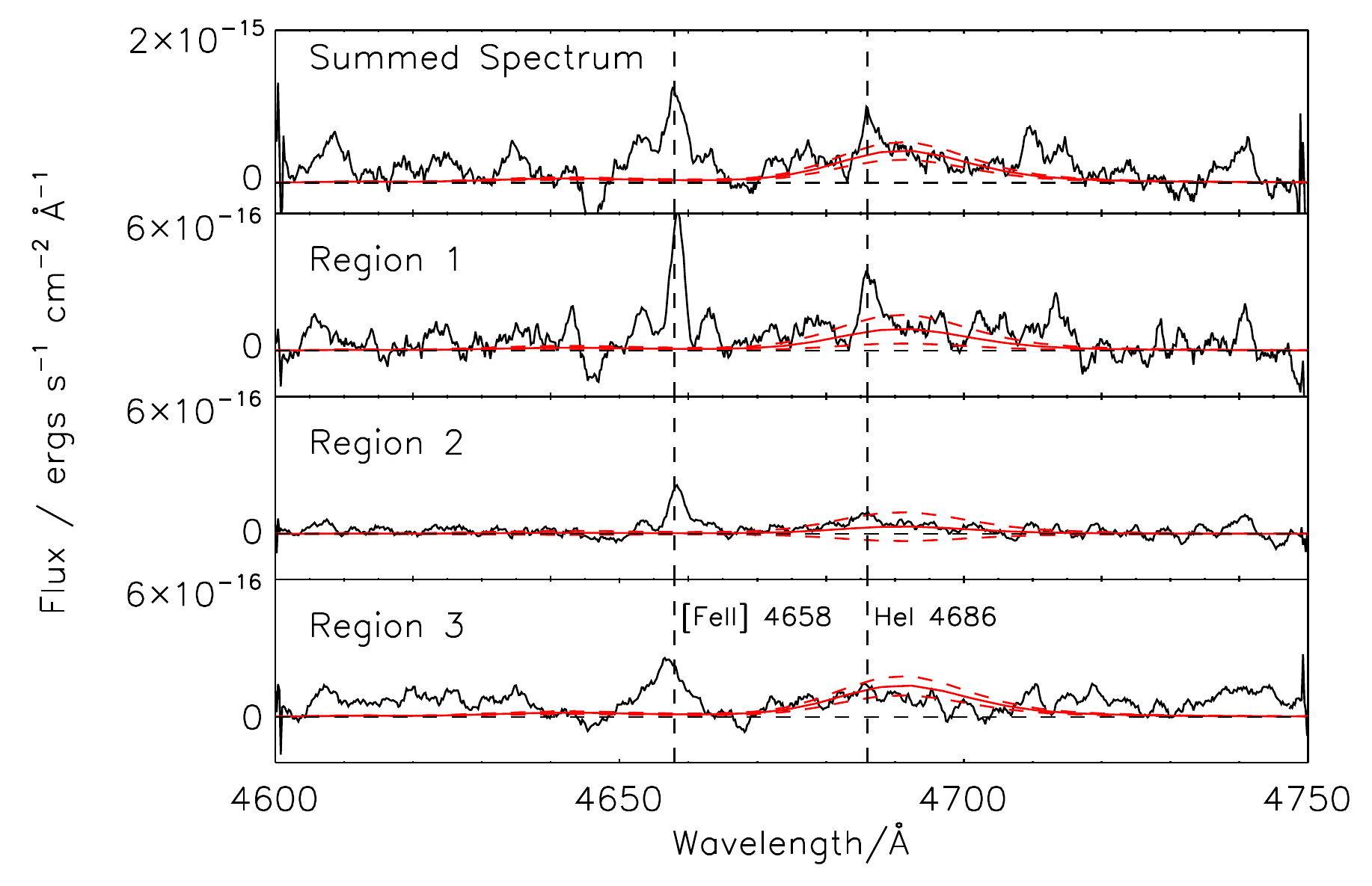}
\caption{Sections of continuum-subtracted FLAMES IFU spectra, showing the blue `WR 
bump' region within summed spectra over the entire 
galaxy and separate aperture regions 1, 2 and 3 (see 
Fig.~\ref{fig:um448_SFR}).  There are minimal detections of WR emission 
features within both the spectra summed over the entire galaxy and also each of the individual aperture region spectra.  Overlaid are the estimates of 4000$\pm$1100, 900$\pm$600, 300$\pm$600 and 1300$\pm$400 WN (5-6) stars (solid red line), for the entire galaxy and regions 1--3, respectively, using the templates of \citet{Crowther:2006}, along with the 1$\sigma$ errors (dashed red line).  Vertical dashed lines indicate the location of [Fe~{\sc iii}]
4658~\AA\, and a weak, narrow, \heii\, emission line at 4686~\AA.  The dashed horizontal lines in the plots show the zero 
flux levels.}
\label{fig:um448_WRspec}
\end{figure} 

Contrary to the conclusions of previous long-slit studies \citep[][and 
references 
therein]{Masegosa:1991,Schaerer:1999} and in agreement with 
\citet{Guseva:2000}, our FLAMES IFU spectra do not reveal evidence of `strong'
WR emission bands at $\sim$4640-4686~\AA.  Unfortunately the 
C~{\sc iv} 5801,5812~\AA\ feature is not included in our wavelength 
coverage, although 
these features are usually significantly weaker than the $\sim$4686~\AA\, 
feature.  We find no evidence of 
a WR population in our single spaxel spectra and only minimal evidence in 
our spectra summed over the entire galaxy. When searching for WR 
features one must consider the width of the extraction aperture, which can 
sometimes be too large and dilute weak WR features by the continuum flux 
\citep{Lopez-Sanchez:2008}.  This has indeed been the case for several other BCGs,
where previous reports of detections of WR features in long-slit spectra 
were not confirmed by IFU spectra \citep[e.g.,][]{James:2010,Perez-Montero:2011}.  We therefore also examined the spectra summed
over each individual region (Figure~\ref{fig:um448_SFR}), which are shown in Fig.~\ref{fig:um448_WRspec}. 

In order to determine the number of WR stars from the strength of the 4640-4690~\AA\ feature for the summed spectra over regions 1-3, the local continuum was subtracted. A linear interpolation to the continuum from both sides of the 4640-4686~\AA\ region was employed based on the mean flux in a blue region (rest wavelengths 4609 -- 4640~\AA, avoiding the \ffeiii\, emission line and possible \ciii\, absorption at 4647~\AA) and a red region (4711 -- 4742~\AA). Utilizing an LMC WR (WN5-6) spectral template from \citet{Crowther:2006}, the best match of the scaled WR star flux to the flux over the observed broad 4686~\AA\, region was determined by minimizing $\chi^{2}$ over the peak of the WR feature, typically 4670 - 4900~\AA. Since errors were available on the flux points these were used to generate a series of Monte Carlo realizations of each region spectrum and the WR template flux was best fitted to each. The distribution the WR template flux to each realization was fitted by a Gaussian enabling the 1$\sigma$ errors to be estimated. The flux of the WR template was then converted into the number of WR stars allowing for the distance of UM~448. We estimate 900 $\pm$ 600, 300 $\pm$ 600 and 1300 $\pm$ 400 WN stars for regions 1-3 respectively. A total population of 4000 $\pm$ 1100 WN stars was estimated applying the same procedure for the summed spectrum for the entire galaxy. These error estimates are based purely on the flux errors and do not take into account any systematic error arising from the, necessarily simplified, removal of a linear continuum under the WR line.The WR template fits and the error ranges for the summed spectra over the entire galaxy and the three regions 1-3 are shown in Fig. \ref{fig:um448_WRspec}.  However, considering the large uncertainties, the existence of a WR population within UM~448 is still considered to be speculative and as a result, we cannot conclude that the $\Delta$log(N/O)$\sim+$0.6~dex excess observed in region~2 is due to nitrogen enrichment from WR stars.


However, following the methodology outlined in \citet{Perez-Montero:2011}, we can use the size of the N-enriched region in UM~448 ($\sim$1.6\arcsec$\times$1.6\arcsec\, = 0.6~kpc$^2$) in region 2 as a scale length to assess whether the reported WR stars could be the actual source of this N-pollution.  Using modeled chemical yields at $\sim$0.2\,\Zsol, those authors study the effects of elements ejected into the ISM via winds as a function of time and pollution radii \citep[see][section 4.3 for a full description of the model parameters]{Perez-Montero:2011}.  After $\sim$4~Myr, the appearance of WR star winds increases the N/O ratio, where the value of N/O ratio only reaches that of region 2 at radii much lower (e.g., $r=10$ or $100$~pc) than those measured from our IFU observations.  Therefore, although WR stars have been detected in previous long-slit studies, and are tentatively detected here, the high N/O ratio observed within region 2 is not likely produced by the pollution from stellar wind ejecta coming from WR stars.  Instead, this enrichment may be related to other global processes occurring within UM~448, namely the likely interaction and/or merger between two bodies as inferred from the SuSI2 images and radial velocity maps.  Studies suggest that interacting galaxies fall $\gtrsim$0.2~dex below the mass-metallicity relation of normal galaxies due to tidally induced large-scale inflow of metal-poor gas to the galaxies' central regions \citep[e.g.][]{Kewley:2006,Michel:2008,Peeples:2009,Rupke:2010,Torrey:2012}.  In addition, models by \citet{Koppen:2005} have shown that a rapid decrease in oxygen abundance during an episode of massive and rapid accretion of metal-poor gas is followed by a slower evolution which leads to the closed-box relation, thus forming a loop in the N/O--O/H diagram.  Such a scenario, in conjunction with metal-rich gas loss driven by SN winds, has been suggested by \citet{Amorin:2010} to explain the systematically large N/O ratio found within \textit{green pea} galaxies.

\section{Summary \& Conclusions}
We have analysed FLAMES-IFU integral field spectroscopy of the BCG UM~448 and studied its
morphology by creating monochromatic emission line maps.  This system shows signs of
interaction and/or perturbation and it is possibly undergoing a merger.  Based on the emission line morphology, we defined three regions of high \ha\, surface brightness which we use to derive error-weighted averages of the physical and chemical conditions across UM~448.  Region 1 is to the east and corresponds to the tidal tail, region 3 extends towards to the upper-west and region 2 lies between them.  UM~448 exhibits complex emission line profiles, with most lines consisting of a narrow, central component and an underlying broad component, and a third, narrow blue-shifted component is also detected in Regions 2 and 3.

Abundance ratio maps of O/H, N/H, Ne/H and S/H were created based on the  direct method of
estimating electron temperatures from the \foiii\, $\lambda$4363/$\lambda$4959 line ratio.
Minimal variance in oxygen abundance are seen across UM~448 for the narrow velocity component, with an average abundance of 12+log(O/H)=8.15$\pm$0.1.  However, in an area of UM~448 where the \foiii\ 4363\,\AA\ line is resolved into a narrow$+$broad component, an $\sim$0.4~dex increase in oxygen abundance attributable to a decrease in \elt\, of $\sim$5000~K is found in the broad line region.  We detect the previously reported high N/O ratio, with $\Delta$log(N/O)$\sim+0.4$, in our aperture region 2 only, spatially coincident with a peak in nitrogen abundance. Abundances derived from an integrated spectrum of UM~448 are not a reliable representation of the ionic and elemental abundances throughout UM~448, which may be a consequence of the variance in \elt\, throughout the nebular emission line components of UM~448 detected in these integral field spectra.  However, abundances relative to oxygen from the two methods (i.e. map-based averages vs. integrated spectra) are consistent within the uncertainties. These results are significant when assessing the reliability of abundances derived from long-slit and integrated spectra of both nearby and high-redshift star-forming galaxies that do not contain homogeneous ionization conditions and/or metallicities.

Continuum images of UM~448 display a rather different
morphology to those of the emission lines, with a peak in stellar light situated in between the second and third
custom-aperture regions. The colour of this continuum peak (from SuSI2 broad-band
images) shows signs of a red, older stellar population in contrast to a younger, bluer
stellar population situated in the eastern body of UM~448. Star formation rates
were computed from the \ha\, line luminosities within each sampled star forming region and the
ages of the last major star formation episode were estimated by fitting the observed
Balmer line equivalent widths with \textsc{starburst99} models. The two merging sections
of UM~448 have SFRs that differ by a factor of $\sim$5 and starburst episodes separated by
4~Myr, providing evidence of a distinct age gradient between the two components.  The latest major
star formation event took place $\sim$4.5~Myr ago in the largest merging component, and
has been producing stars at a rate of 0.8~\Msol\, yr$^{-1}$.

Finally, we do not detect the previously reported `strong' WR signatures in UM~448, on a single-spaxel, regional or global scale \citep{Masegosa:1991,Schaerer:1999}.  Using the 4686~\AA\, feature we tentatively estimate that $\sim$4000 WN stars may be present within UM~448. In consideration of the physical size of increased N/O seen within UM~448 ($\sim$0.6~kpc$^2$) we find that WR stars alone cannot be responsible for the region of N-enrichment within UM~448.  Thus being, another chemical evolution scenario (e.g. inflow of metal-poor gas due to interaction or merger scenario, as the morphology and dynamics of UM~448 suggest) is required to explain this enhancement.  Since there is no strong evidence in support of these scenarios (although broad component emission is suggestive of an outflow, the N-enrichment is seen in the narrow component only), more observations, at higher sensitivity, are needed of this intriguing system.   In relation to other high N/O BCGs, this chemodynamical study supports its membership within the anomalous subset and, along with other spatially resolved spectroscopic studies of high N/O galaxies, demonstrates the benefits that this method can have in both isolating the enrichment and aiding our understanding of the processes that cause it.

\bibliographystyle{mn2e}
\bibliography{references.bib}

\section{acknowledgments}
We would like to thank the FLAMES staff at Paranal and Garching  for scheduling and taking
these service mode observations [programme 083.B-0336B; PI: B. L. James].  We appreciate
discussions with Alessandra Aloisi regarding theories of primary/secondary nitrogen
production. We also thank the referee for the useful comments that have significantly improved the first submitted version of this manuscript.  YGT acknowledges the award of a Marie Curie intra-European
Fellowship within the 7th European Community Framework Programme
(grant agreement PIEF-GA-2009-236486). BLJ acknowledges support from an STScI-DDRF travel grant used to visit collaborators of this study.  This research made use of the NASA ADS and NED data bases.


\appendix
\section{Regional Observed Fluxes and Line Intensities}
Tables~\ref{tab:um448-1}--\ref{tab:um448-3} list the observed fluxes and de-reddened line intensities derived from summed spectra over the regions defined in Fig.~\ref{fig:um448_SFR}.  Fluxes are given for each separate velocity component (C1--C3), along with its de-reddened line intensity and FWHM.
\begin{table*}
\begin{center}
\begin{scriptsize}
\begin{tabular}{l|r@{$\pm$}lr@{$\pm$}lr@{$\pm$}lr@{$\pm$}lr@{$\pm$}lr@{$\pm$}lr@{$\pm$}lr@{$\pm$}lr@{$\pm$}lr@{$\pm$}lr@{$\pm$}lr@{$\pm$}l}
\hline

	& \multicolumn{6}{c|}{C1}														& \multicolumn{6}{c|}{C2}											\\
\hline																													
	&	\multicolumn{2}{c}{F$_\lambda$	}		&	\multicolumn{2}{c}{I$_\lambda$}			&	\multicolumn{2}{c}{FHWM} 			&	\multicolumn{2}{c}{F$_\lambda$}			&	\multicolumn{2}{c}{I$_\lambda$}			&	\multicolumn{2}{c}{FHWM}\\
\foii\, $\lambda$3727	&	93.39	&	1.84	&	108.46	&	3.62	&	77.07	&	0.48	&	69.57	&	2.33	&	100.49	&	9.06	&	167.96	&	1.65	\\
\foii\, $\lambda$3729	&	110.36	&	2.38	&	128.17	&	4.43	&	77.02	&	0.48	&	129.04	&	3.49	&	186.39	&	16.39	&	167.81	&	1.64	\\
\fneiii\, $\lambda$3868	&	8.69	&	0.35	&	9.94	&	0.48	&	58.50	&	14.57	&	35.36	&	3.74	&	49.28	&	6.60	&	169.82	&	15.62	\\
H8+\hei\, $\lambda$3888	&	10.69	&	6.14	&	12.22	&	7.02	&	70.76	&	17.14	&	25.16	&	8.63	&	34.91	&	12.31	&	172.18	&	38.13	\\
\hd\, 	&	18.53	&	1.38	&	20.61	&	1.63	&	80.58	&	2.20	&	32.73	&	2.02	&	42.51	&	4.25	&	145.87	&	2.70	\\
\hg\,	&	43.78	&	2.10	&	47.16	&	2.54	&	86.13	&	1.85	&	40.17	&	2.72	&	48.21	&	4.88	&	175.53	&	4.32	\\
\foiii\, $\lambda$4363	&	2.93	&	0.27	&	3.15	&	0.30	&	104.09	&	10.38				\\
\hei\, $\lambda$4471	&	3.40	&	1.01	&	3.60	&	1.07	&	74.58	&	14.14	&	5.02	&	1.27	&	5.76	&	1.52	&	236.59	&	52.09	\\
\hb\,	&	100.00	&	2.23	&	100.00	&	3.09	&	81.32	&	0.69	&	100.00	&	3.01	&	100.00	&	7.31	&	189.73	&	2.19	\\
\foiii\, $\lambda$4959	&	91.61	&	2.46	&	90.34	&	3.08	&	76.61	&	0.82	&	100.14	&	3.49	&	96.75	&	7.14	&	182.57	&	2.65	\\
\fnii\, $\lambda$6548	&	15.15	&	0.57	&	12.58	&	0.51	&	69.33	&	1.26	&	14.51	&	0.76	&	9.19	&	0.64	&	156.65	&	4.10	\\
\ha\,	&	349.49	&	18.66	&	291.23	&	16.13	&	104.33	&	0.35	&	467.65	&	10.91	&	298.72	&	15.36	&	235.81	&	1.00	\\
\fnii\ $\lambda$6584	&	56.17	&	1.15	&	46.50	&	1.17	&	77.19	&	0.65	&	37.07	&	1.15	&	23.31	&	1.28	&	194.92	&	3.40	\\
\hei\, $\lambda$6678	&	0.50	&	0.16	&	0.41	&	0.14	&	30.78	&	6.20	&	9.16	&	0.29	&	5.65	&	0.31	&	87.88	&	1.67	\\
\fsii\, $\lambda$6716	&	37.41	&	0.82	&	30.63	&	0.80	&	71.74	&	0.55	&	40.41	&	1.11	&	24.73	&	1.28	&	155.42	&	1.32	\\
\fsii\, $\lambda$6731	&	29.24	&	0.63	&	23.95	&	0.62	&	71.59	&	0.55	&	0.91	&	0.02	&	0.56	&	0.03	&	155.08	&	1.32	\\
																									
c(\hb) &\multicolumn{6}{c|}{		0.25	$\pm$	0.01	}		&\multicolumn{6}{c|}{						0.62	$\pm$	0.03		}							\\
F(\hb) &\multicolumn{6}{c|}{		5.30	$\pm$	0.08	}		&\multicolumn{6}{c|}{						3.83	$\pm$	0.08		}							\\

\hline
\end{tabular}

\caption{UM~448 regional fluxes and de-reddenened line intensities (both relative to F(\hb)=I(\hb)=100) and FWHMs (in \kms) for summed spectra over region 1 (as defined in Fig.~\ref{fig:um448_SFR}).  Line fluxes were extinction-corrected using the c(\hb) values shown at the bottom of the table, calculated from the relative \ha, \hb\, and \hg\, fluxes.  F(\hb) is in units of $\times10^{-14}~ergs~s^{-1}~cm^{-2}~arcsec^{-2}$.} 
\label{tab:um448-1}
\end{scriptsize}
\end{center}
\end{table*}

\begin{table*}
\begin{center}
\begin{scriptsize}
\begin{tabular}{l|r@{$\pm$}lr@{$\pm$}lr@{$\pm$}lr@{$\pm$}lr@{$\pm$}lr@{$\pm$}lr@{$\pm$}lr@{$\pm$}lr@{$\pm$}lr@{$\pm$}lr@{$\pm$}lr@{$\pm$}l}
\hline

	& \multicolumn{6}{c|}{C1}														& \multicolumn{6}{c|}{C2}											\\
\hline																													
	&	\multicolumn{2}{c}{F$_\lambda$	}		&	\multicolumn{2}{c}{I$_\lambda$}			&	\multicolumn{2}{c}{FHWM} 			&	\multicolumn{2}{c}{F$_\lambda$}			&	\multicolumn{2}{c}{I$_\lambda$}			&	\multicolumn{2}{c}{FHWM}\\
\foii\, $\lambda$3727	&	135.44	&	4.93	&	165.56	&	11.27	&	88.05	&	1.05	&	73.89	&	3.39	&	97.79	&	7.94	&	197.23	&	3.84	\\
\foii\, $\lambda$3729	&	160.52	&	6.28	&	196.21	&	13.65	&	87.98	&	1.05	&	117.41	&	4.18	&	155.40	&	11.78	&	197.06	&	3.84	\\
\fneiii\, $\lambda$3868	&	30.52	&	5.22	&	36.59	&	6.59	&	102.24	&	10.23	&	14.84	&	3.93	&	19.11	&	5.21	&	323.00	&	92.41	\\
H8+\hei\, $\lambda$3888	&	8.18	&	7.09	&	9.78	&	8.49	&	60.54	&	25.61	&	25.41	&	5.01	&	32.61	&	6.77	&	152.59	&	21.87	\\
\hd\, 	&	14.24	&	3.86	&	16.43	&	4.54	&	80.17	&	7.25	&	27.99	&	2.90	&	34.16	&	4.14	&	139.71	&	3.95	\\
\hg\,	&	42.40	&	3.00	&	46.84	&	4.10	&	94.70	&	4.38	&	41.56	&	2.73	&	47.76	&	4.25	&	170.55	&	7.05	\\
\foiii\, $\lambda$4363	&	3.10	&	0.54	&	3.41	&	0.62	&	124.72	&	24.87	\\
\hei\, $\lambda$4471	&	1.43	&	1.89	&	1.54	&	2.04	&	56.25	&	34.72	&	4.60	&	1.30	&	5.11	&	1.47	&	140.16	&	29.63	\\
\hb\,	&	100.00	&	3.78	&	100.00	&	5.94	&	91.60	&	1.35	&	100.00	&	3.80	&	100.00	&	6.55	&	200.23	&	3.45	\\
\foiii\, $\lambda$4959	&	102.17	&	3.89	&	100.26	&	5.89	&	86.57	&	1.55	&	67.05	&	3.16	&	65.31	&	4.59	&	191.52	&	4.09	\\
\fnii\, $\lambda$6548	&	23.57	&	1.59	&	18.37	&	1.36	&	81.05	&	2.43	&	12.88	&	1.07	&	9.09	&	0.82	&	165.02	&	5.90	\\
\ha\,	&	370.44	&	11.04	&	290.03	&	12.58	&	103.56	&	0.57	&	415.93	&	12.45	&	295.55	&	13.99	&	278.17	&	2.12	\\
\fnii\ $\lambda$6584	&	97.70	&	2.92	&	75.84	&	3.27	&	92.68	&	0.88	&	31.35	&	1.13	&	22.01	&	1.12	&	253.20	&	4.52	\\
\hei\, $\lambda$6678	&	9.41	&	0.33	&	7.23	&	0.34	&	102.59	&	2.28		\\
\fsii\, $\lambda$6716	&	53.03	&	2.02	&	40.55	&	1.97	&	85.70	&	1.06	&	29.47	&	1.24	&	20.26	&	1.11	&	180.53	&	2.77	\\
\fsii\, $\lambda$6731	&	43.70	&	1.59	&	33.42	&	1.58	&	85.52	&	1.06	&	18.79	&	0.91	&	12.92	&	0.77	&	180.14	&	2.76	\\
																									
c(\hb) &\multicolumn{6}{c|}{		0.34	$\pm$	0.02	}		&\multicolumn{6}{c|}{						0.48	$\pm$	0.02		}							\\
F(\hb) &\multicolumn{6}{c|}{		1.95	$\pm$	0.05	}		&\multicolumn{6}{c|}{						2.72	$\pm$	0.07		}							\\

 \hline
\end{tabular}

\caption{Same as for Table~\ref{tab:um448-1}, for summed spectra over region 2 (as defined in Fig.~\ref{fig:um448_SFR}).}
\label{tab:um448-2}
\end{scriptsize}
\end{center}
\end{table*}

\begin{table*}
\begin{scriptsize}
\begin{tabular}{l|r@{$\pm$}lr@{$\pm$}lr@{$\pm$}lr@{$\pm$}lr@{$\pm$}lr@{$\pm$}lr@{$\pm$}lr@{$\pm$}lr@{$\pm$}lr@{$\pm$}lr@{$\pm$}lr@{$\pm$}l}
\hline	
	& \multicolumn{6}{c|}{C1}														& \multicolumn{6}{c|}{C2}											\\
\hline																													
	&	\multicolumn{2}{c}{F$_\lambda$	}		&	\multicolumn{2}{c}{I$_\lambda$}			&	\multicolumn{2}{c}{FHWM} 			&	\multicolumn{2}{c}{F$_\lambda$}			&	\multicolumn{2}{c}{I$_\lambda$}			&	\multicolumn{2}{c}{FHWM}\\
\foii\, $\lambda$3727	&	376.20	&	7.41	&	506.95	&	52.62	&	199.37	&	0.73													\\
\foii\, $\lambda$3729	&	487.54	&	9.55	&	656.99	&	68.17	&	212.46	&	15.14													\\
\fneiii\, $\lambda$3868	&	65.57	&	4.66	&	85.82	&	10.52	&	213.38	&	19.76													\\
H8+\hei\, $\lambda$3888	&	55.24	&	5.09	&	72.05	&	9.78	&	221.94	&	2.30													\\
\hd\, 	&	9.51	&	2.11	&	11.76	&	2.85	&	95.75	&	9.88	&	26.11	&	1.09	&	32.67	&	1.87	&	157.59	&	5.69	\\
\hg\,	&	39.89	&	3.33	&	46.25	&	5.73	&	113.23	&	26.08	&	38.99	&	1.16	&	45.59	&	2.18	&	220.76	&	22.26	\\
\foiii\, $\lambda$4363	&	1.45	&	0.84	&	1.67	&	0.97	&	74.34	&	37.30	&	5.33	&	0.07	&	6.19	&	0.24	&	455.65	&	67.72	\\
\hei\, $\lambda$4471	&	13.64	&	1.33	&	15.25	&	2.01	&	226.35	&	22.75													\\
\hb\,	&	100.00	&	2.71	&	100.00	&	8.56	&	118.95	&	8.93	&	100.00	&	0.96	&	100.00	&	3.46	&	233.89	&	7.09	\\
\foiii\, $\lambda$4959	&	97.79	&	2.17	&	95.09	&	7.82	&	94.79	&	3.52	&	89.70	&	0.97	&	87.09	&	2.98	&	255.45	&	4.99	\\
\fnii\, $\lambda$6548	&	60.50	&	2.56	&	41.76	&	2.91	&	188.97	&	2.54													\\
\ha\,	&	414.10	&	8.58	&	287.88	&	17.14	&	111.69	&	4.90	&	417.95	&	3.94	&	284.58	&	7.03	&	249.28	&	5.71	\\
\fnii\ $\lambda$6584	&	132.16	&	2.76	&	90.71	&	5.33	&	162.09	&	7.08	&	35.66	&	3.51	&	23.95	&	2.42	&	333.30	&	17.45	\\
\hei\, $\lambda$6678	&	14.28	&	0.71	&	9.66	&	0.71	&	200.80	&	7.71													\\
\fsii\, $\lambda$6716	&	176.68	&	4.77	&	118.61	&	7.10	&	193.47	&	0.69													\\
\fsii\, $\lambda$6731	&	135.17	&	4.06	&	90.75	&	5.56	&	193.01	&	0.69													\\
																									
c(\hb) &\multicolumn{6}{c|}{		0.51	$\pm$	0.04	}		&\multicolumn{6}{c|}{						0.53	$\pm$	0.01		}							\\
F(\hb) &\multicolumn{6}{c|}{		1.67	$\pm$	0.03	}		&\multicolumn{6}{c|}{						3.54	$\pm$	0.02		}							\\
\hline

\end{tabular}
\begin{tabular}{l|r@{$\pm$}lr@{$\pm$}lr@{$\pm$}lr@{$\pm$}lr@{$\pm$}lr@{$\pm$}l}

\hline	
	& \multicolumn{6}{c|}{C3}								\\
\hline																													
	&	\multicolumn{2}{c}{F$_\lambda$	}		&	\multicolumn{2}{c}{I$_\lambda$}			&	\multicolumn{2}{c}{FHWM} \\
\foii\, $\lambda$3727													\\
\foii\, $\lambda$3729													\\
\fneiii\, $\lambda$3868													\\
H8+\hei\, $\lambda$3888													\\
\hd\, 													\\
\hg\,	&	41.93	&	3.60	&	47.58	&	5.45	&	106.30	&	20.01	\\
\foiii\, $\lambda$4363													\\
\hei\, $\lambda$4471													\\
\hb\,	&	100.00	&	5.80	&	100.00	&	8.88	&	84.49	&	6.54	\\
\foiii\, $\lambda$4959	&	129.25	&	8.12	&	126.20	&	11.47	&	86.50	&	4.99	\\
\fnii\, $\lambda$6548													\\
\ha\,	&	401.60	&	16.70	&	294.40	&	18.32	&	86.10	&	4.71	\\
\fnii\ $\lambda$6584	&	77.15	&	3.50	&	55.94	&	3.60	&	91.24	&	6.78	\\
\hei\, $\lambda$6678													\\
\fsii\, $\lambda$6716													\\
\fsii\, $\lambda$6731													\\
													
c(\hb) &\multicolumn{6}{c|}{		0.43	$\pm$	0.03		}							\\
F(\hb) &\multicolumn{6}{c|}{		0.54	$\pm$	0.02		}							\\
\hline
\end{tabular}
\caption{Same as for Table~\ref{tab:um448-1}, for summed spectra over region 3 (as defined in Fig.~\ref{fig:um448_SFR}).}
\label{tab:um448-3}
\end{scriptsize}
\end{table*}

\bsp

\label{lastpage}

\end{document}